\documentclass[11pt]{article}
\pdfoutput=1
\usepackage{amsmath, amssymb}
\usepackage{bm}
\usepackage{graphicx}
\usepackage{wrapfig}
\usepackage[T1]{fontenc}

\begin{document}
\title{\bf Multiplicity distribution in pseudo-rapidity windows and charge conservation} 

\author{  Naomichi Suzuki$^{1}$,
             Minoru Biyajima$^{2}$ 
             and  Takuya Mizoguchi$^{3}$
             \thanks{\emph{e-mail:} mizoguti@toba-cmt.ac.jp}\\
{ $^{1}$Matsumoto University, Matsumoto 390-1295, Japan} \\
{ $^{2}$Department of Physics, Shinshu University, Matsumoto 390-8621, Japan} \\ 
{ $^{3}$National Institute of Technology, Toba College, Toba 517-8501, Japan} 
}
 
\date{}
\maketitle 
\begin{abstract}
 Charged multiplicity distribution in a pseudo-rapidity window is formulated under 
the assumption that the charge conservation is satisfied in the full phase space. 
 At first, we analyze measured charged particle multiplicity distributions in pseudo-rapidity windows in LHC by the CMS and ALICE collaborations with the two probability distributions. One is the convolution of negative binomial and Poisson distributions, and the other is the Glauber-Lachs formula. 
  Each distribution is considered as an analogy of the quantum optics. 
Next, we analyze the data with the double GL formulae for $|\eta|<2.4$ 
at 7 TeV by the CMS collaboration and for $|\eta|<1.5$ at 8 TeV by the ALICE 
collaboration to describe the global structure of measured distributions. 
\end{abstract}
\section{Introduction}
%%-----------------

In the middle of 1980's, multiplicity distributions of charged particles in pseudo-rapidity windows were reported in the CERN $p\bar{p}$ collider 
experiments~\cite{Alner1984}. 
To analyze the data, a multiplicity distribution which is a convolution of a negative  
binomial distribution (NBD) and a Poisson distribution 
 (PSND) was proposed~\cite{Fowler1986}:     
\begin{eqnarray}
  &&P(n,\langle n\rangle) 
          = \sum_{n=n_1+n_2}P(n_1,\langle n_1\rangle)P_2(n_2,\langle n_2\rangle),
                \label{eq.int01} \\
  && P_1(n_1,\langle n_1\rangle) = \frac{ (n_1+k-1)! }{ n_1!(k-1)! } 
               \frac{(\langle n_1\rangle/k)^{n_1}}{ (1+\langle n_1\rangle/k)^{n_1+k} },  
                \label{eq.int02}\\
  && P_2(n_2,\langle n_2\rangle) 
               = \frac{ \langle n_2 \rangle^{n_2} }{ n_2! } e^{-\langle n_2 \rangle}. 
                  \label{eq.int03} 
\end{eqnarray}
In the above equations, $\langle n\rangle$, $\langle n_1\rangle$, and $\langle n_2 \rangle$ denote the average multiplicities in each distribution, where a relation, $\langle n\rangle = \langle n_1\rangle + \langle n_2 \rangle $, holds.  
Three parameters, $k$, $\langle n\rangle$ and $\tilde{p} = \langle n_1\rangle/\langle n \rangle$ are contained in Eq.(\ref{eq.int01}).
The NBD corresponds to the distribution of particles emitted from the chaotic sources in the thermal equilibrium, and the PSND to that of particles emitted from the coherent source.
After the analysis of measured negative charged multiplicity distributions in the full phase space at $\sqrt{s}=540$ GeV by the use of 
Eq.(\ref{eq.int01}) with $k=1$ and $k=2$,  measured charged multiplicity distributions in the pseudo-rapidity windows were analyzed to estimate the value of $\tilde{p}$.  
A stochastic background of Eq.(\ref{eq.int01}) was investigated in \cite{Biyajima1991}.

In a model of identical particle correlations based on the quantum optical approach~\cite{Glauber1966,Lachs1965}, particles emitted from chaotic sources 
and those from coherent source are correlated~\cite{Biyajima1978,Suzuki1997,Csorgo1999}. Therefore, 
the multiplicity distribution composed of chaotic and coherent components is not necessarily written by two independent distributions such as Eq. (\ref{eq.int01}).

In \cite{Suzuki2011}, a multiplicity distribution obtained from semi-inclusive momentum distributions in the quantum optical approach  
 have been presented:
\begin{eqnarray}
  &&P(n,\langle n\rangle) 
           = \frac{ (p_{\rm in}\langle n \rangle)^n }{(1+p_{\rm in}\langle n\rangle)^{n+1}}
     \exp\Bigl[ - \frac{ (1-p_{\rm in})\langle n\rangle }{  1 + p_{\rm in}\langle n\rangle } \Bigr]
     L_n\Bigl( -\frac{ (1-p_{\rm in})/p_{\rm in} } { 1 + p_{\rm in}\langle n\rangle } \Bigr),   \label{eq.int05}  
\end{eqnarray}
where, $p_{\rm in}$ denotes the ratio of the average multiplicity of negative charged particles emitted from the chaotic source to $\langle n \rangle$. 
Equation (\ref{eq.int05}) is called Glauber-Lachs (GL) formula~\cite{Glauber1966, Lachs1965, Biyajima1984}

In the LHC experiments, charged particle multiplicity distributions are measured in restricted pseudo-rapidity windows. In the full phase space, charge conservation 
should be satisfied. 
Therefore, we would like to consider a relation between the charged multiplicity distribution in a pseudo-rapidity window and that in the full phase space. In addition,  we would like to investigate 
some characteristics in  $\tilde{p}$ and $p_{\rm in}$ analyzing the measured  multiplicity distributions in the recent LHC experiments by Eq.(\ref{eq.int01}) with 
$k=1$, or $2$, and Eq.(\ref{eq.int05}).

In \cite{Pumplin1994}, measured charged multiplicity distributions at 
$\sqrt{s} = 1800$ GeV  were analysed by the use of the QCD Monte Carlo program 
 HERWIG (versin 5.7). Due to the contribution of two jets events with large pseudorapidity gap, a relation on the mupliplicity distribution, $P(0) \simeq 2P(1)$, 
  was obtained. The result is similar to measured charged multiplicity distributions 
  in the LHC experiments.
 
In the invariant energy $\sqrt{s}$ above several hundred GeV, it is considered that it would be very hard to describe measured multiplicity distributions with a single probability distribution~\cite{Giovannini1999,Dremin2004,Ghosh2012,Zborovsky2013,ALICE2017,Rybczynski2018, Biyajima2018}.  We also try to fit the data with double GL formulae.

The present paper is organized as follows. In section 2, charged multiplicity distribution in a pseudo-rapidity window is formulated under the assumption that the charge conservation is satisfied in the full phase space. In section 3, charged multiplicity distributions in pseudo-rapidity windows measured in the LHC experiments are analyzed by the use of Eq.(\ref{eq.int01}) and Eq.(\ref{eq.int05}).  
Moreover, double GL formulae are used in the analysis.
Section 4 is devoted to concluding remarks.  
Detail calculations for some equations in section 2, and explicit expressions of charged multiplicity distributions for PSND, NBD and generalized Glauber-Lachs 
 (GGL)  formula  in the pseudo-rapidity window are shown in appendix \ref{apdx.mdcc}.

For comparison, we also analyze the data, directly using Eq.(\ref{eq.int01}) with two parameters $\tilde{p}$ and $\langle n\rangle$. The results are shown 
in appendix \ref{apdx.FW}.   
In appendix  \ref{apdx.GL}, Data are also analyzed by Eq.(\ref{eq.int05}) with two 
 parameters, $p_{\rm in}$ and $\langle n_{\rm ch} \rangle$ 
in place of  $\langle n \rangle$. 

%%-------------------
\section{Charged multiplicity distribution in a pseudo-rapidity window with charge conservation in the full phase space}
%%------------------------------------------------------------------

In the full phase space, the measured multiplicity distribution satisfies the charge conservation. For simplicity, we assume that the charged particles are produced in pairs of a positive charged particle and a negative charged particle. Let $P(n)$, $n=0,1,\ldots$ be a multiplicity distribution of negative charged particles, and 
$P_{\rm ch}(2n)$ be that of charged particles in the full phase space. We assume that a relation, 
 \begin{eqnarray}
    P(n)=P_{\rm ch}(2n) \label{eq.tl01}, 
 \end{eqnarray}
holds. 

Furthermore, we would like to adopt the following assumption: A probability that each particle produced in the full phase space enters into a limited window (and is detected) is $\zeta$ ( $0\le \zeta \le 1$ ), and that each particle does not enter into the window is $1-\zeta$. 
When more than $n$ pairs of charged particles are produced in the full phase space, 
and $m$ ($2n\le m\le 0$) charged particles enter into the pseudo-rapidity window, the probability distribution  $P_{\rm ob}(m)$ that $m$ charged particles enter into the window is written as,
 \begin{eqnarray}
    P_{\rm ob}(m) = \sum_{2n\ge m}^\infty {}_{2n}C_m 
                  \zeta^m(1-\zeta)^{2n-m} P_{\rm ch}(2n).        \label{eq.tl02}
 \end{eqnarray}

In the following, $P(n)$ is written as $P(n,\langle n\rangle)$ with the average multiplicity $\langle n\rangle$ of negative charged particles in the full phase space.
A multiplicity distribution $P_\zeta(j)$ is defined as 
 \begin{eqnarray}
      P_\zeta (j) \equiv \sum_{n=j}^{\infty} {}_nC_j\,  
           \bigl[ \zeta(2-\zeta) \bigr]^j 
           \bigl[ (1-\zeta)^2 \bigr]^{n-j} P(n,\langle n\rangle), 
            \quad j=0,1,2,\cdots,                 \label{eq.tl03} 
 \end{eqnarray}
which denotes the multiplicity distribution that when $n$ pairs  ($n\ge j$) of charged 
particles are produced, $(n-j)$ pairs are outside the pseudo-rapidity window, 
and at least one particle enters into the window from any $j$ pairs of negative and 
 positive charged particles. 
 
Relations among $P_{\rm ob}(n)$, $P_\zeta(j)$ and $P(n,\langle n\rangle)$ are shown in Appendix \ref{apdx.mdcc}. 
We obtain from Eq.(\ref{eq.ap_ob06}):
 \begin{eqnarray}
       P_{\rm ob}(n)  = \sum_{j=0}^{[n/2]} {}_{n-j} C_{j}\,
           \frac{ (\zeta^2)^{j} [ 2\zeta(1-\zeta) ]^{n-2j} }{ [ \zeta(2-\zeta) ]^{n-j} } 
                               P_\zeta(n-j).    \label{eq.tl04}          
 \end{eqnarray}

In the present paper, we use three distribution functions, PSND, NBD and GL formula for $P(n,\langle n \rangle)$. In any of the three distribution functions, the following relation holds:
 \begin{eqnarray}
       P_\zeta(n) = P(n,\langle n_\zeta \rangle), \quad 
           \langle n_\zeta \rangle = \zeta(2-\zeta) \langle n \rangle.  \label{eq.tl05} 
 \end{eqnarray}
From Eqs.(\ref{eq.tl04}) and (\ref{eq.tl05}),  the multiplicity distribution $P_{\rm ob}(n)$
 of charged particles in the pseudo-rapidity window is expressed with that $P(n,\langle n\rangle)$ of negative charged particles in the full phase space as, 
 \begin{eqnarray}
     &&  P_{\rm ob}(n)  = \sum_{j=0}^{[n/2]} {}_{n-j} C_{j}\,
           \frac{ (\zeta^2)^{j} [ 2\zeta(1-\zeta) ]^{n-2j} }{ [ \zeta(2-\zeta) ]^{n-j} } 
                               P(n-j,\langle n_\zeta \rangle).    \label{eq.tl06}          
 \end{eqnarray}

If we can omit $\zeta^2$ which is regarded as much smaller than $\zeta$ in 
Eq.(\ref{eq.tl06}), we obtain a relation,
 \begin{eqnarray}
       P_{\rm ob}(n)  \simeq P(n, 2\zeta \langle n \rangle).    \label{eq.tl07}          
 \end{eqnarray}
% 

%%-------------
\section{Analysis of charged multiplicity distributions in pseudo-rapidity windows}
%%------------------------------------------------------------------

At first, the invariant energy $\sqrt{s}$ dependence of average charged multiplicity, $\langle n_{\rm ch}\rangle$, in the full phase space in non-single diffractive (NSD)  events is parametrized as,
 \begin{eqnarray}
    \langle n_{\rm ch} \rangle = 0.986{s}^{1/4} + 6.309,   \label{eq.anal01} 
 \end{eqnarray}
by the least mean square method with the data from $\sqrt{s}=30.4$ GeV to $\sqrt{s}=1800$ GeV~\cite{Break1984, Ansor1989, Alexo1998}. 
The average multiplicity of negative charged particles in the full phase space, 
$\langle n \rangle$, is estimated from Eq.(\ref{eq.anal01}) with the relation 
$\langle n \rangle = \langle n_{\rm ch} \rangle/2$.  
Those used in the present analysis~\cite{ALICE2017,CMS2011} 
are listed in Table \ref{tab.avmlt}.

 \begin{table}[htb]
  \caption{ Average multiplicities of negative charged particles in the full phase space  used in the analysis. }
  \label{tab.avmlt}
 \begin{center}
  \begin{tabular}{cccccc} \hline \hline
   $\sqrt{s}$ (TeV) & 0.9 &  2.36 & 2.76 & 7  & 8  \\  
  \hline 
    $\langle n \rangle = \langle n_{\rm ch} \rangle/2$ & 17.9 & 27.1 & 29.1 & 44.4 & 47.3 \\ 
  \hline  \hline  
  \end{tabular}
 \end{center}
 \end{table}

In the experiments of Bose-Einstein correlations (BEC), the number of identical boson pairs, say $\pi^{-}$ pairs $N^{(2-)}$ relative to the number of uncorrelated pion pairs  $N^{BG}$ as a function of relative momentum squared, $Q=\sqrt{-(p_1-p_2)^2}$, is measured, and for example, it is fitted by 
 \begin{eqnarray*}
     N^{(2-)}/N^{BG} = C[ 1 + \lambda\, \Omega(Qr) ](1 + \delta\, Q).     
 \end{eqnarray*}
Function $\Omega(Qr)$ is often parametrized as $\Omega(Qr) = e^{-Qr}$.  Normalization factor $C$ is determined so as to $ N^{(2-)}/N^{BG} \simeq 1$ for $Q>>1$.

In the quantum optical approach to the BEC~\cite{Biyajima1980}, the second order BEC function is given by 
 \begin{eqnarray*}
   N^{(2-)}/N^{BG} = 1 + 2p_{\rm in}( 1-p_{\rm in} ) e^{-Qr} + {p_{\rm in}}^2 e^{-2Qr}.     
 \end{eqnarray*}
Therefore, the following relation is satisfied: 
 \begin{eqnarray}
   \lambda = p_{\rm in}( 2-p_{\rm in} ).    \label{eq.anal02}
 \end{eqnarray}

In the present analysis, we estimate the value of $p_{\rm in}$ from the measured charged multiplicity distribution. Data samples used in the BEC experiments are different from those used in the charged multiplicity measurements~\cite{CMS2010,ATLAS2015}. For example, in the CMS collaboration, BEC data are taken for $p_T > 200$ MeV and $|\eta|<2.4$~\cite{CMS2010}.  
On the other hand, measured charged multiplicity distributions in pseudo-rapidity windows are taken for $p_T > 0$ MeV. Therefore, it is not clear whether Eq.(\ref{eq.anal02}) is satisfied or not.  We would like to compare the estimated value of 
$ p_{\rm in}( 2-p_{\rm in} ) $ with parameter $\lambda$ estimated from the BEC 
 experiments.

%--------------------
\subsection{Analysis with Eq.(\ref{eq.tl06}) and the convolution of NBD and PSND}
%-------------------------------------------------------------------

We analyze the charged multiplicity distributions of non-single diffractive (NSD) events in the pseudo-rapidity window, $|\eta|<\Delta\eta$~\cite{CMS2011, ALICE2017}. 

At first, we analyze measured charged multiplicity distributions by the CMS Collaboration in pseudo-rapidity windows, $|\eta|<\Delta\eta$, at $\Delta\eta=0.5, 1,0,1.5,2.0$ and $2.4$ with Eq.(\ref{eq.tl06}) and the convolution of NBD and PSND 
 given by the following equation with $k=1$ or $2$,  
\begin{eqnarray}
  &&P(n,\langle n_\zeta\rangle) = \sum_{n=n_1+n_2} \frac{ (n_1+k-1)! }{ n_1!(k-1)! } 
      \frac{(\langle n_{1\zeta}\rangle/k)^{n_1}}{ (1+\langle n_{1\zeta}\rangle/k)^{n_1+k} } 
  \times\frac{ \langle n_{2\zeta} \rangle^{n_2} }{n_2!} e^{-\langle n_{2\zeta} \rangle}, 
    \label{eq.anal03} \\
   && \hspace{15mm}  \langle n_{1\zeta}\rangle = \tilde{p}\zeta(2-\zeta)\langle n\rangle, \quad 
      \langle n_{2\zeta}\rangle = (1-\tilde{p})\zeta(2-\zeta) \langle n\rangle. \nonumber 
\end{eqnarray}

\begin{figure}[htb]
  \centering
     \includegraphics[width=6.8cm,clip]{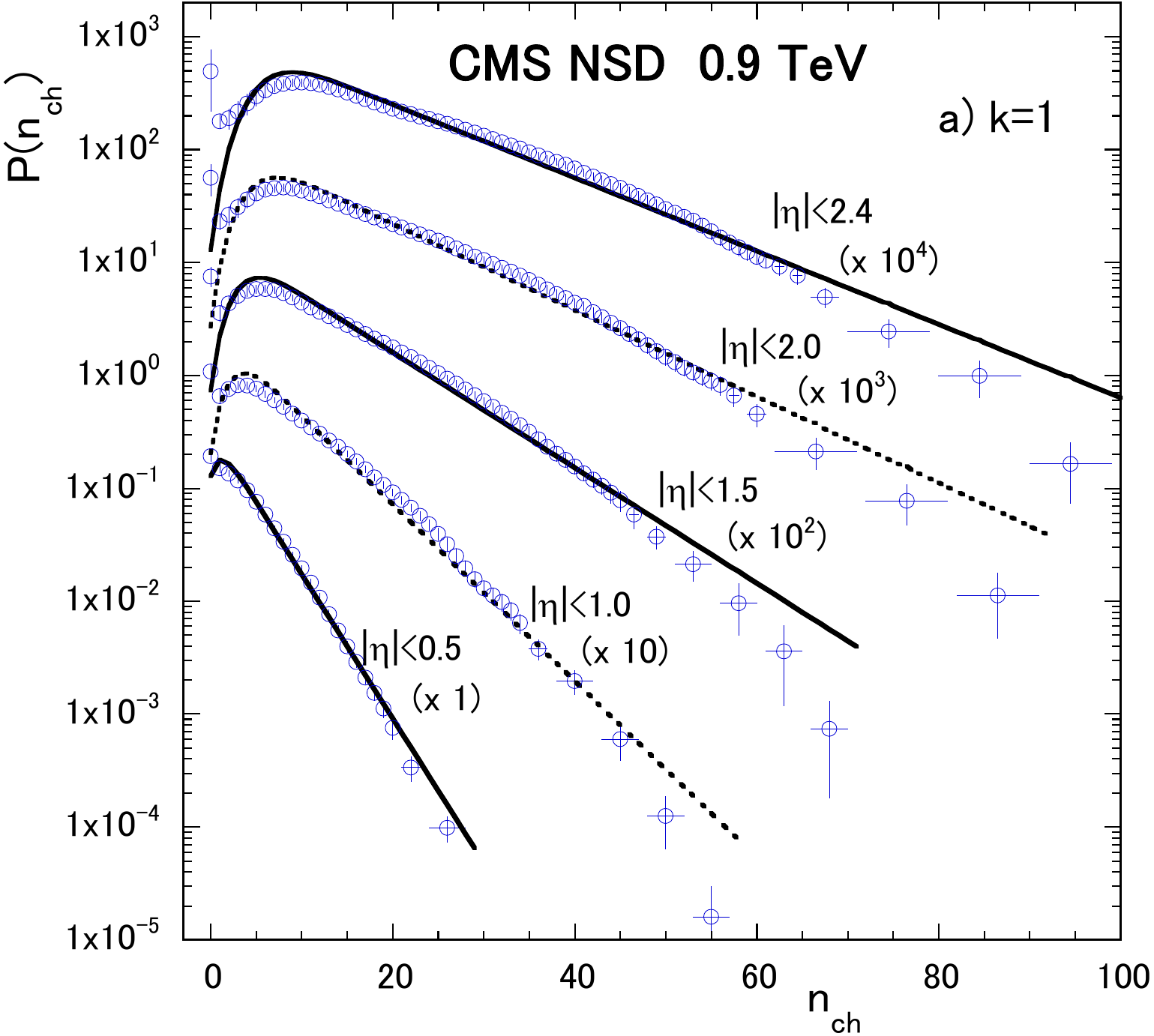}
    \hspace{3mm}
     \includegraphics[width=6.8cm,clip]{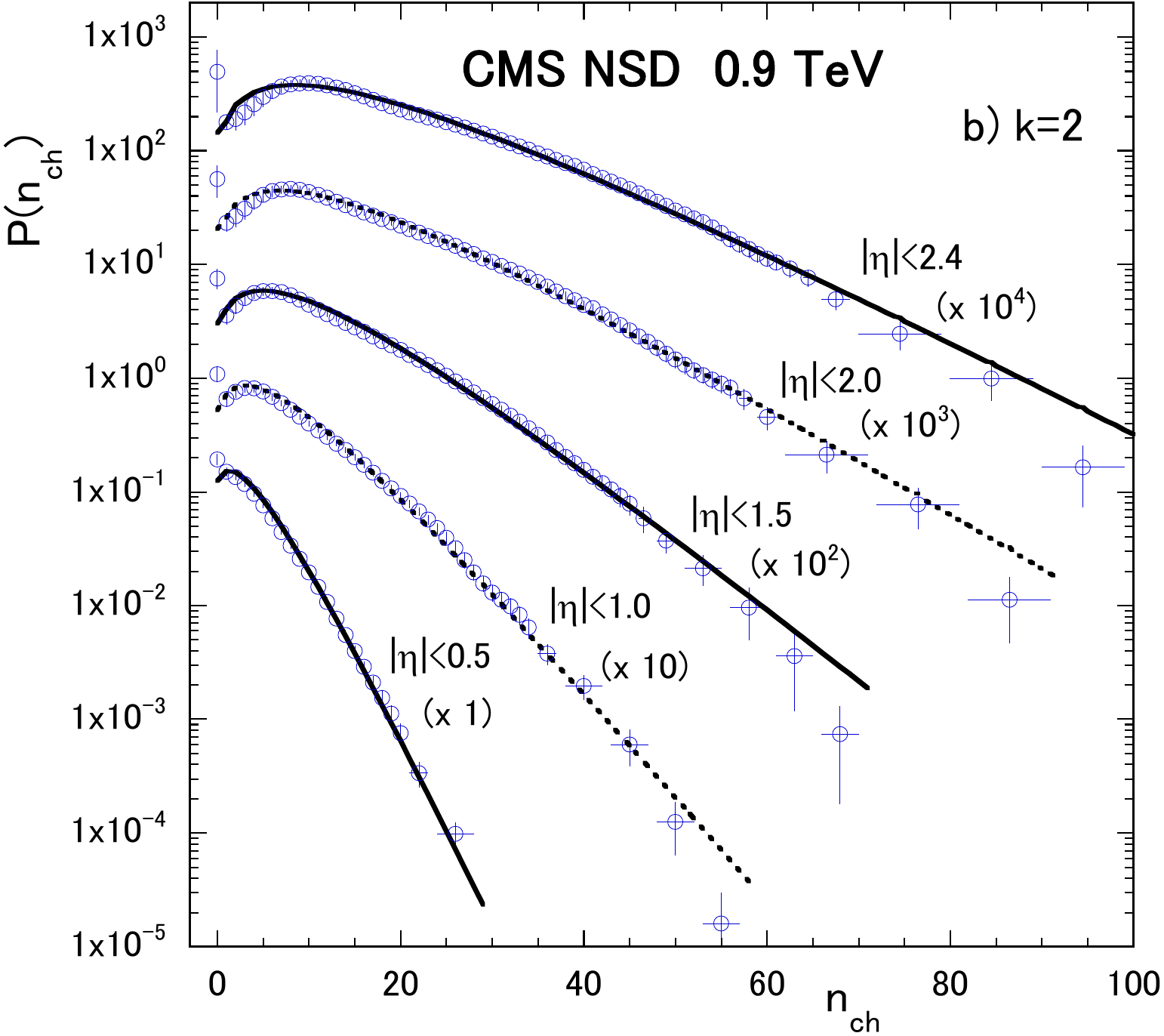}
  \caption{Charged multiplicity distributions at $\sqrt{s}=0.9$ TeV compared to  theoretical curves (solid or dotted lines) calculated with Eqs.(\ref{eq.tl06}) and (\ref{eq.anal03}) : a) $k=1$ and b) $k=2$. }
  \label{fig.cms0900FW} 
\end{figure}

Results on the charged multiplicity distributions at $\sqrt{s}=0.9$ TeV by the CMS Collaboration by Eqs.(\ref{eq.tl06}) and (\ref{eq.anal03}) 
are shown in Fig.\ref{fig.cms0900FW} and Table \ref{tab.cms0900FW}. 

%---------- cms0900_FW ----------
%
 \begin{table}[ht]
  \caption{Parameters estimated from the analysis of charged multiplicity 
 distributions at $\sqrt{s}=0.9$ TeV by the CMS Collaboration 
 by Eqs.(\ref{eq.tl06}) and (\ref{eq.anal03}) with $k=1$ or $k=2$. }
  \label{tab.cms0900FW}
  {\small
 \begin{center}
  \begin{tabular}{ccccccc} \hline \hline
  $\sqrt{s}$ (TeV) &  $k$ & $\Delta \eta$ &  $\tilde{p}$  & $\zeta$ 
                                  & $\chi_{\rm min}^2/{\rm n.d.f.}$ 
                                  & $\langle n\rangle_{\rm ob}=2\zeta\langle n\rangle$  \\ 
   \hline
   $0.9$&$1$& $0.5$ & $0.787\pm 0.028$  & $0.102\pm 0.002$ & $63.7/(23-2)$
                                                        & $3.65 \pm 0.08$ \\ 
          &     & $1.0$ & $0.670\pm 0.025$  & $0.207\pm 0.005$ & $299.2/(40-2)$ 
                                                        & $7.41 \pm 0.18$ \\     
          &     & $1.5$ & $0.696\pm 0.019$  & $0.315\pm 0.006$ & $263.2/(52-2)$ 
                                                        & $11.3 \pm 0.2$ \\     
          &     & $2.0$ & $0.695\pm 0.015$  & $0.428\pm 0.006$ & $249.5/(62-2)$ 
                                                        & $15.3 \pm 0.2$ \\     
          &     & $2.4$ & $0.685\pm 0.015$  & $0.517\pm 0.008$ & $298.4/(68-2)$ 
                                                        & $18.4\pm 0.3$ \\ 
  \hline    
   $0.9$&$2$&$0.5$ & $ 1.000\pm 0.035$  & $0.108\pm 0.002$ & $51.1/(23-2)$  
                                                        & $3.87\pm 0.08$ \\ 
          &     &$1.0$ & $1.000\pm 0.024$   & $0.210\pm 0.003$ & $85.0/(40-2)$ 
                                                        & $7.52\pm 0.11$ \\     
          &     &$1.5$ & $1.000\pm 0.017$   & $0.316\pm 0.003$ & $69.6/(52-2)$ 
                                                        & $11.3\pm 0.1$ \\     
          &     &$2.0$ & $1.000\pm 0.012$   & $0.423\pm 0.003$ & $49.2/(62-2)$ 
                                                        & $15.1\pm 0.1$ \\     
          &     &$2.4$ & $0.990\pm 0.011$ & $0.507\pm 0.004$ & $64.3/(68-2)$ 
                                                        & $18.2\pm 0.1$ \\ 
  \hline \hline
   \end{tabular}  
 \end{center}
 }
 \end{table}

At $\sqrt{s}=0.9$ TeV, the results with $k=2$ describes the data better than 
those with $k=1$. 
%The minimum $\chi^2$ value in each analysis with $k=2$ is greater than 1 except for  that for $|\eta|<2.0$.
%
In this case, the estimated value of $\tilde{p}$ with $k=1$ is almost 1. 
Therefore, the coherent component in multiplicity distribution is almost 0 and the chaotic component is to occupy almost 100 percent of multiplicities 
at $\sqrt{s}=0.9$ TeV.

%------------------------
%-------------------------

%---------- cms2360_FW_fig ----------
 \begin{figure}[htb]
  \centering
    \includegraphics[width=6.8cm,clip]{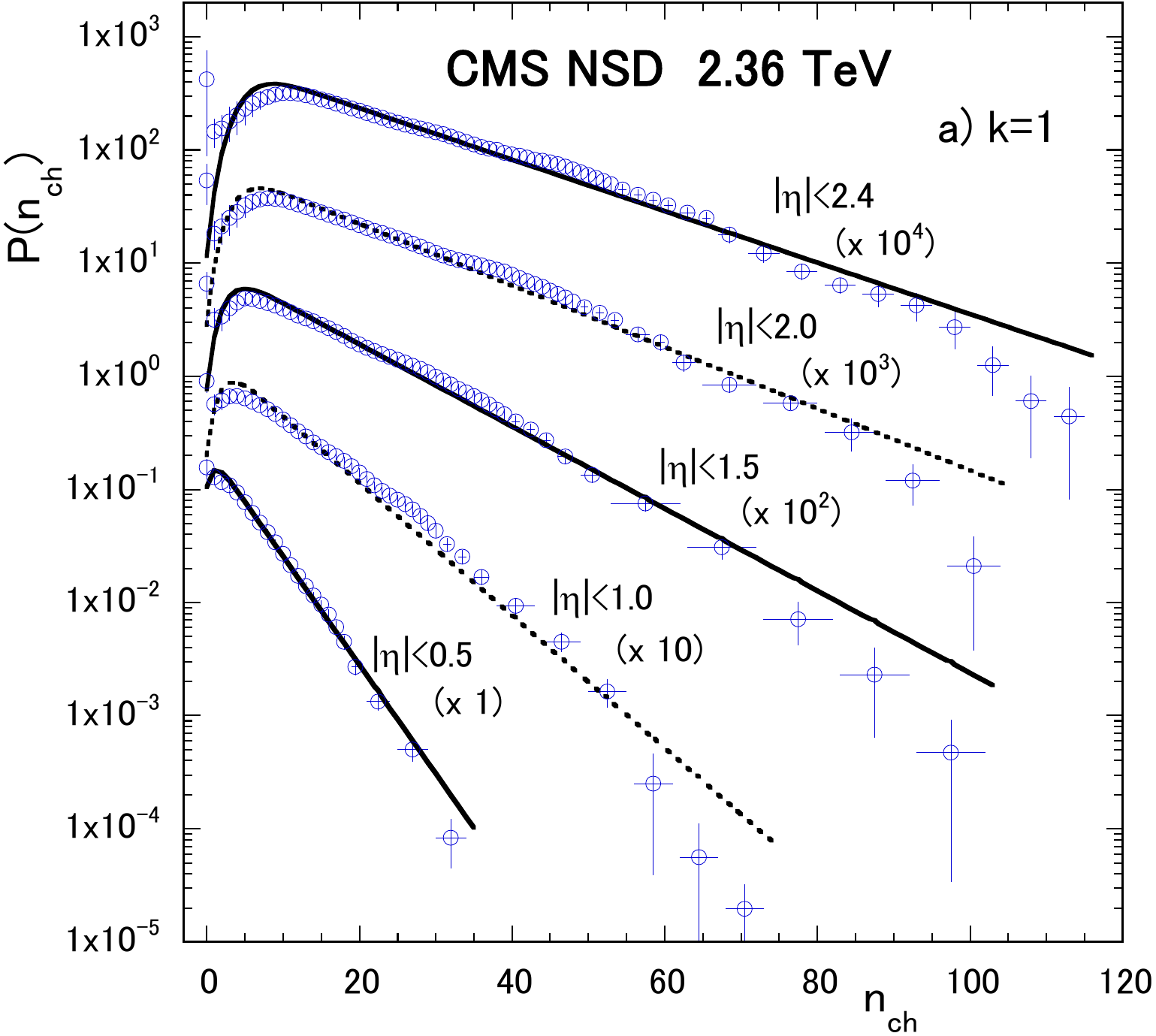}
   \hspace{3mm}
     \includegraphics[width=6.8cm,clip]{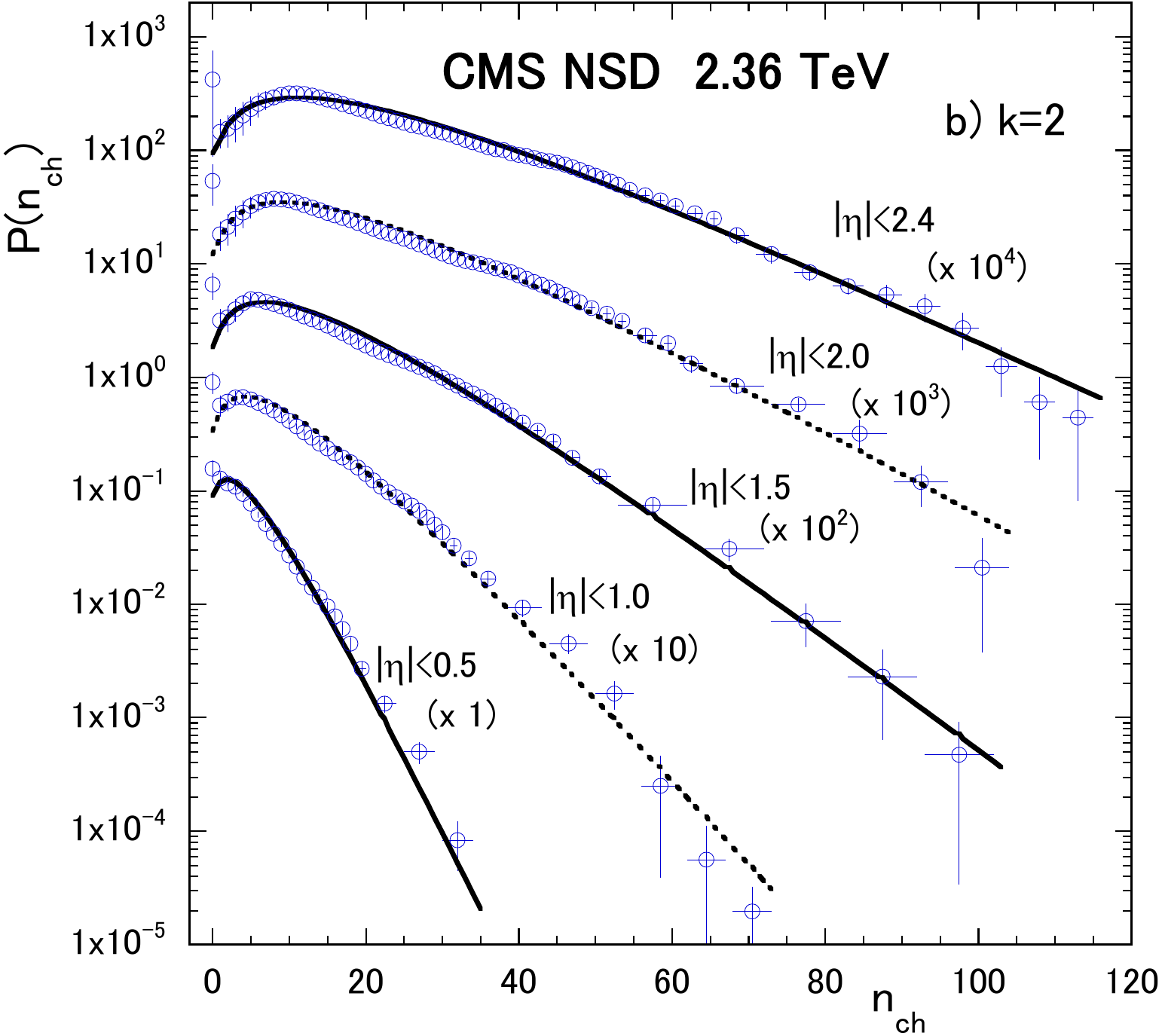}
   \caption{Charged multiplicity distributions at $\sqrt{s}=2.36$ TeV  compared to 
 theoretical curves (solid or dotted lines)  calculated with  Eqs.(\ref{eq.tl06}) 
and (\ref{eq.anal03}) : a) $k=1$ and b) $k=2$. }
  \label{fig.cms2360FW} 
\end{figure}  
%
%---------- cms2360_FW_tab ----------
%
 \begin{table}[ht]
  \caption{Parameters estimated from the analysis of charged multiplicity 
 distributions at $2.36$ TeV by the CMS Collaboration 
 by Eqs.(\ref{eq.tl06}) and (\ref{eq.anal03}) with $k=1$ and $k=2$. }
  \label{tab.cms2360FW}
  {\small
 \begin{center}
  \begin{tabular}{ccccccc} \hline \hline
  $\sqrt{s}$ (TeV) &  $k$ & $\Delta \eta$ &  $\tilde{p}$  & $\zeta$ 
                                  & $\chi_{\rm min}^2/{\rm n.d.f.}$ 
                                  & $\langle n\rangle_{\rm ob}=2\zeta\langle n\rangle$ \\ 
   \hline
 $2.36$&$1$&$0.5$& $0.846\pm 0.022$  & $0.0871\pm 0.0016$ & $41.4/(23-2)$  
                                                      & $4.72\pm 0.09$  \\ 
          &     &$1.0$& $0.765\pm 0.028$  & $0.165\pm 0.005$ & $315.4/(40-2)$ 
                                                      & $8.94\pm 0.26$ \\     
          &     &$1.5$& $0.798\pm 0.017$  & $0.262\pm 0.005$ & $152.3/(50-2)$ 
                                                      & $14.2\pm 0.3$  \\     
          &     &$2.0$& $0.794\pm 0.015$  & $0.355\pm 0.006$ & $155.1/(60-2)$ 
                                                      & $19.2\pm 0.3$ \\     
          &     &$2.4$& $0.781\pm 0.013$  & $0.434\pm 0.007$ & $143.1/(70-2)$ 
                                                      & $23.5\pm 0.4$ \\ 
  \hline    
 $2.36$&$2$&$0.5$& $1.000\pm 0.047$  & $0.090\pm 0.002$ & $105.0/(23-2)$ 
                                                      & $4.89\pm 0.13$ \\ 
        &     &$1.0$&  $1.000\pm  0.031$  & $0.180\pm 0.004$ & $152.3/(40-2)$ 
                                                      & $9.76\pm 0.20$ \\     
        &     &$1.5$&  $1.000\pm  0.024$  & $0.273\pm 0.005$ & $128.3/(50-2)$ 
                                                      & $14.8 \pm 0.2$ \\     
        &     &$2.0$&  $1.000\pm  0.020$  & $0.364\pm 0.005$ & $115.0/(60-2)$ 
                                                      & $19.7\pm 0.3$ \\     
        &     &$2.4$&  $1.000\pm  0.016$  & $0.438\pm 0.005$ & $82.2/(70-2)$ 
                                                      & $23.7\pm 0.3$ \\ 
  \hline \hline
   \end{tabular}  
 \end{center}
 }
 \end{table}

Results at $\sqrt{s}=2.36$ TeV are shown in Fig.\ref{fig.cms2360FW} and 
Table \ref{tab.cms2360FW}.  
At $\sqrt{s}=2.36$ TeV, the results with $k=2$ describes the data better than 
those with $k=1$ except for the data for $|\eta|<0.5$.  
The value of $\chi^2_{\rm min}$/n.d.f in each analysis with $k=2$ is greater than 1, 
and estimated values of $\tilde{p}$ become almost 1. 
That for $|\eta|<0.5$ with $k=1$ is 1.65.  

At $\sqrt{s}=7$ TeV, the results with $k=1$ and with $k=2$ can not fit the data well.

\vspace{4mm}

For comparison, we also analyze the data, directly using Eq.(\ref{eq.int01}) with two parameters $\tilde{p}$ and $\langle n_{\rm ch} \rangle$. $\langle n \rangle$ in Eq.(\ref{eq.int01}) is replaced by $\langle n_{\rm ch}  \rangle$. 
Results at $\sqrt{s}=0.9$ and 2.36 TeV are shown respectively 
in Tables \ref{tab.cms0900FWb} and \ref{tab.cms2360FWb} in appendix \ref{apdx.FW}.

%---------
\subsection{Analysis with Eq.(\ref{eq.tl06}) and the GL formula}
%---------------------------------------------------

\begin{figure}[ht]
  \centering
    \includegraphics[width=6.8cm,clip]{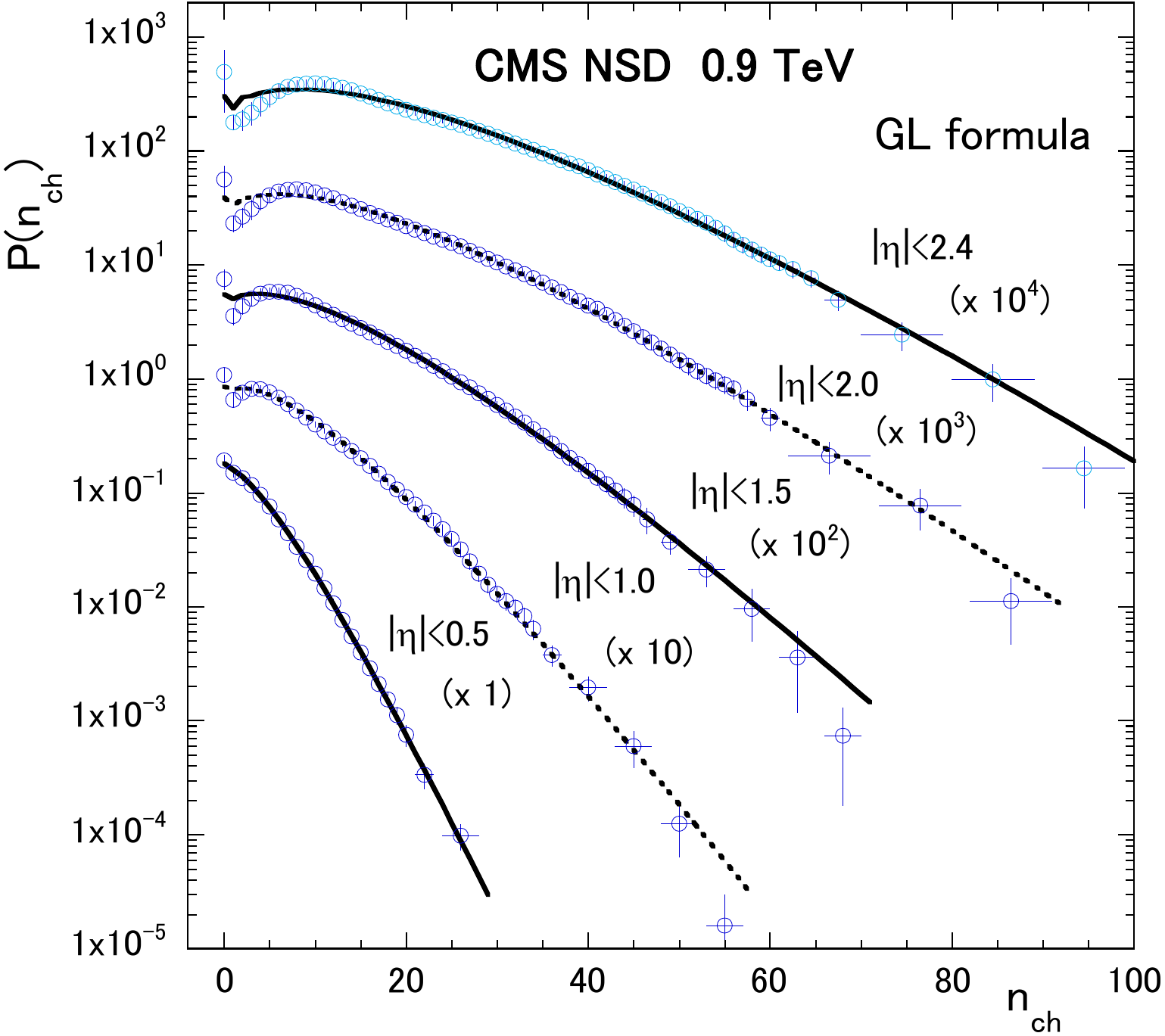}
    \hspace{3mm}
    \includegraphics[width=6.8cm,clip]{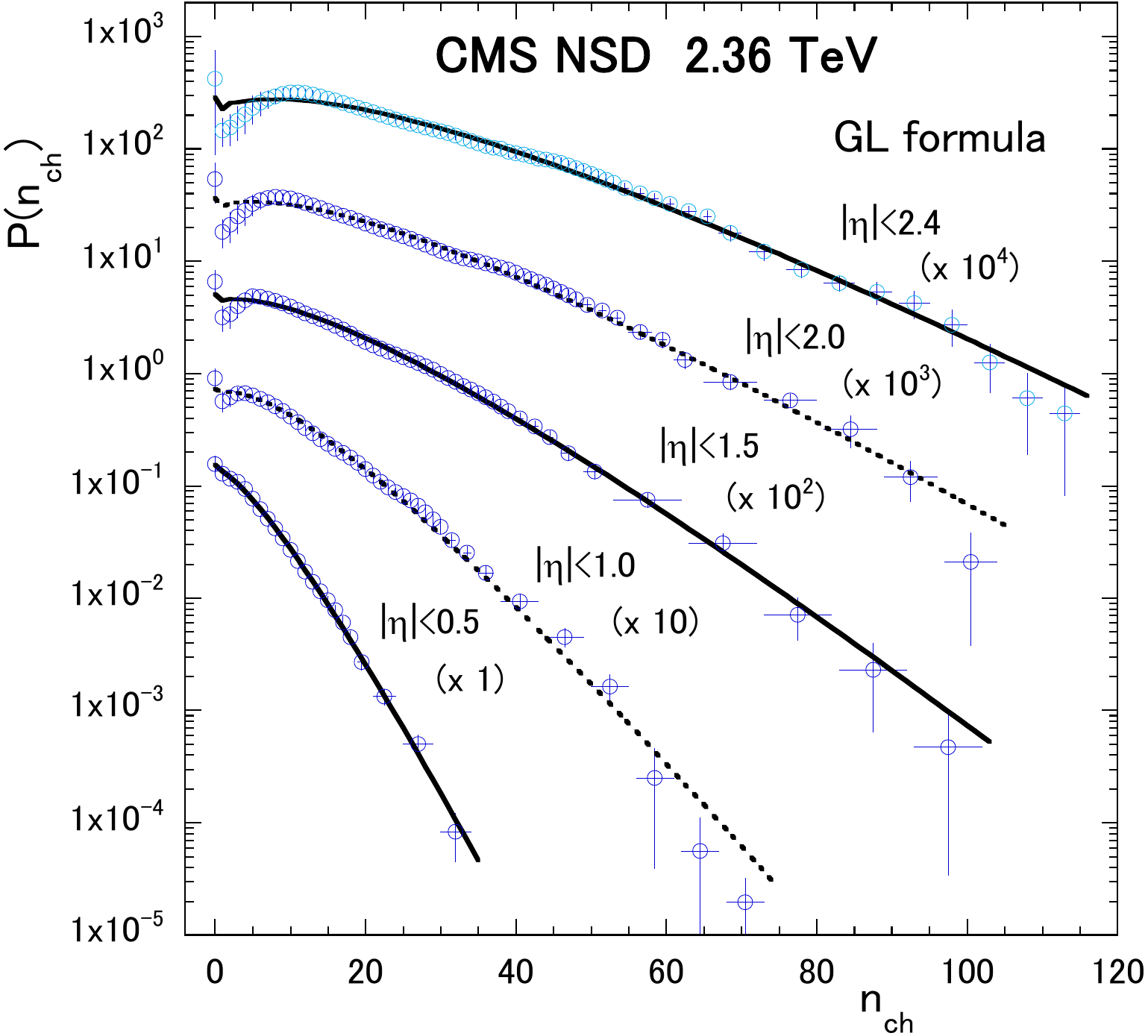}
  \\
    \includegraphics[width=6.8cm,clip]{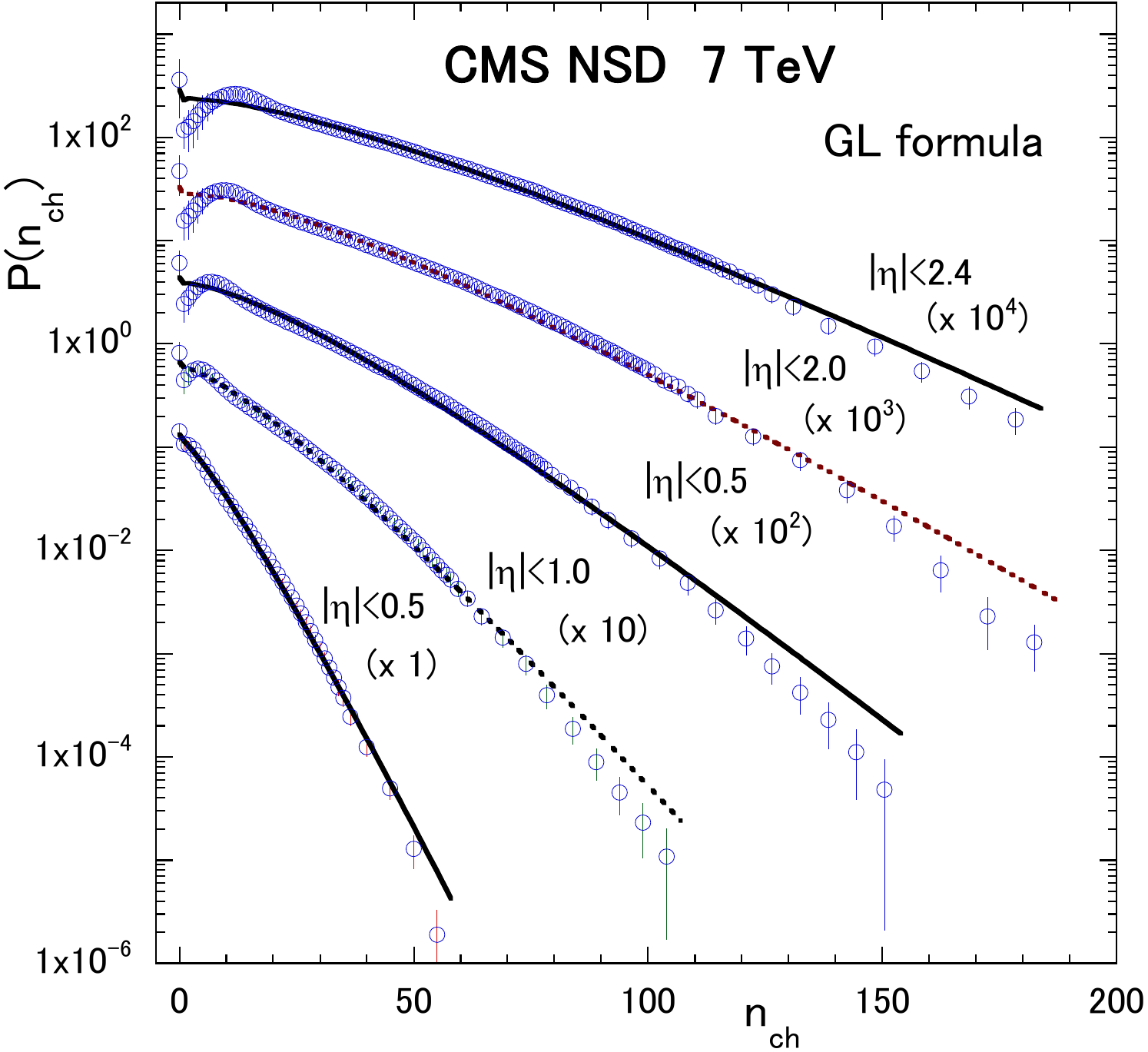}
  \caption{Charged multiplicity distributions at $\sqrt{s}=0.9$, 2.36 and 7 TeV 
compared to theoretical curves (solid or dotted lines) 
calculated with Eqs.(\ref{eq.tl06}) and (\ref{eq.anal04}). }
  \label{fig.cms_GL1} 
\end{figure}

Next, we would like to analyze measured charged multiplicity distributions 
with Eq.(\ref{eq.tl06}) and the GL formula, 
 \begin{eqnarray}
      P(n,\langle n_\zeta \rangle) 
         = \frac{(p_{\rm in} \langle n_\zeta \rangle)^n}
               {(1+p_{\rm in}\langle n_\zeta \rangle)^{n+1}} 
     \exp\Bigl[
    -\frac{(1-p_{\rm in}) \langle n_\zeta \rangle }{1+p_{\rm in} \langle n_\zeta \rangle} 
      \Bigr]
    L_n\Bigl( -\frac{(1-p_{\rm in})/ p_{\rm in} }{1+p_{\rm in} \langle n_\zeta \rangle}  \Bigr),  \label{eq.anal04} 
 \end{eqnarray}
where, $\langle n_\zeta \rangle = \zeta(2-\zeta)\langle n \rangle$.
Results on the charged multiplicity distributions at $\sqrt{s}=0.9$,  
$2.36$ and $7$ TeV by the CMS Collaboration by Eqs.(\ref{eq.tl06}) 
and (\ref{eq.anal04}),  
are shown in Fig.\ref{fig.cms_GL1} and Table \ref{tab.cms_GL}.

%----------- CMS Fig -----------------
%-----------------

%---------- CMS Tab -----------
%
 \begin{table}[htb]
  \caption{Parameters estimated from the analysis of charged multiplicity distributions at $\sqrt{s}=0.9,  2.36$ and $7$ TeV 
by the CMS collaboration by  Eqs.(\ref{eq.tl06}) and (\ref{eq.anal04}). }
  \label{tab.cms_GL}
 \begin{center}
 {\small
  \begin{tabular}{ccccccc} \hline \hline
   $\sqrt{s}$ (TeV) & $\Delta \eta$ &  $p_{\rm in}$  & $\zeta$ 
                             & $\chi_{\rm min}^2/{\rm n.d.f.}$ & $p_{\rm in}(2-p_{\rm in})$ 
                             & $\langle n\rangle_{\rm ob}=2\zeta\langle n\rangle$ \\  
  \hline
    $0.9$ & $0.5$ & $0.421\pm 0.007$ & $0.101\pm 0.000$ & $2.2/(23-2)$ 
                                                   & $0.668\pm 0.004$ & $3.62\pm 0.02$ \\ 
            & $1.0$ & $0.358\pm 0.010$  & $0.203\pm 0.002$ & $34.3/(40-2)$ 
                                                   & $0.590\pm 0.006$ & $7.27\pm 0.07$ \\     
            & $1.5$ & $0.342\pm 0.008$  & $0.307\pm 0.002$ & $36.5/(52-2)$  
                                                   & $0.568\pm 0.005$ & $11.0\pm 0.1$ \\     
            & $2.0$ & $0.319\pm 0.007$  & $0.415\pm 0.003$ & $41.8/(62-2)$ 
                                                   & $0.539\pm 0.005$ & $14.9\pm 0.1$  \\     
            & $2.4$ & $0.298\pm 0.007$  & $0.501\pm 0.003$ & $57.7/(68-2)$ 
                                                   & $0.510\pm 0.005$ & $17.9\pm 0.1$ \\ 
  \hline    
   $2.36$& $0.5$ & $0.482\pm 0.014$  & $0.0850\pm 0.0006$ & $ 7.4/(23-2)$  
                                                   & $0.750\pm 0.008$ & $4.61\pm 0.03$ \\ 
            & $1.0$ & $0.406\pm 0.014$  & $0.170\pm 0.002$ & $44.5/(40-2)$ 
                                                   & $0.658\pm 0.009$ & $9.21\pm 0.11$  \\     
            & $1.5$ & $0.416\pm 0.010$  & $0.259\pm 0.002$ & $25.6/(50-2)$ 
                                                   & $0.666\pm 0.006$ & $14.0\pm 0.1$  \\     
            & $2.0$ & $0.386\pm 0.012$  & $0.348\pm 0.003$ & $50.0/(60-2)$ 
                                                   & $0.628\pm 0.007$ & $18.9\pm 0.2$ \\     
            & $2.4$ & $0.360\pm 0.011$  & $0.421\pm 0.004$ & $53.9/(70-2)$ 
                                                   & $0.589\pm 0.008$ & $22.8\pm 0.21$  \\ 
   \hline    
    $7$   & $0.5$ & $0.588\pm 0.020$ & $0.0672\pm 0.0006$ & $ 83.0/(41-2)$ 
                                                   & $0.838\pm 0.009$ & $5.97\pm 0.06$  \\ 
           & $1.0$ & $0.528\pm 0.011$  & $0.137\pm 0.001$ & $130.8/(70-2)$ 
                                                   & $0.779\pm 0.006$ & $12.2\pm 0.08$ \\     
           & $1.5$ & $0.494\pm 0.010$  & $0.207\pm 0.001$ & $191.8/(95-2)$ 
                                                   & $0.747\pm 0.005$ & $18.4\pm 0.12$  \\     
           & $2.0$ & $0.477\pm 0.008$  & $0.280\pm 0.002$ & $203.6/(115-2)$ 
                                                   & $0.731\pm 0.005$ & $24.9\pm 0.1$ \\     
           & $2.4$ & $0.488\pm 0.007$  & $0.340\pm 0.001$ & $133.8/(127-2)$ 
                                                   & $0.741\pm 0.004$ & $30.2\pm 0.1$ \\ 
  \hline \hline   
  \end{tabular}
  }
 \end{center}
 \end{table}

At $\sqrt{s}=0.9$ TeV, values of $\chi^2_{\rm min}/{\rm n.d.f.}$ are less than 1 in all pseudo-rapidity windows. 
At $2.36$ TeV, values of $\chi^2_{\rm min}/{\rm n.d.f.}$ are less than 1 
except for 1.16 for $|\eta|<1.0$.
 At $\sqrt{s}=7$ TeV, values of $\chi^2_{\rm min}/{\rm n.d.f.}$ are less than 2 except for 2.01 for $|\eta|<0.5$. 
 As can be seen from the Tables \ref{tab.cms0900FW}, \ref{tab.cms2360FW} 
and \ref{tab.cms_GL}, results with Eq.(\ref{eq.tl06}) and the GL formula, 
Eq.(\ref{eq.anal04}), describe the data better than those with Eqs.(\ref{eq.tl06}) 
and (\ref{eq.anal03}) for  all pseudo-rapidity windows at $\sqrt{s}=0.9$ and $2.36$ TeV 
by the CMS Collaboration.

%------------------------------------------------------------------ 
Measured values of parameter $\lambda$ are $\lambda = 0.616 \pm 0.031$ 
at $\sqrt{s} = 0.9$ TeV, $\lambda = 0.663 \pm 0.087$ at $\sqrt{s} = 2.36$ TeV,
 and   $\lambda = 0.618 \pm 0.043$ at $\sqrt{s} = 7$ TeV 
by the CMS Collaboration~\cite{CMS2010}. 
By the ATLAS Collaboration~\cite{ATLAS2015}, $\lambda = 0.74 \pm 0.11$ 
at $\sqrt{s} = 0.9$ TeV, and $\lambda = 0.71 \pm 0.07$ at $\sqrt{s} = 7$ TeV.
 
Values of $p_{\rm in}(2-p_{\rm in})$ estimated from the analysis of charged multiplicity 
distributions at $\sqrt{s}=0.9$ TeV are smaller than $\lambda=0.616$ except for 0.668 
at $|\eta|<0.5$.  
Estimated values of $p_{\rm in}(2-p_{\rm in})$ at $\sqrt{s}=2.36$ TeV are not larger 
 than $\lambda=0.663 + 0.087$ for all pseudo-rapidity windows. 
Estimated values of $p_{\rm in}(2-p_{\rm in})$ at $\sqrt{s}=7$ TeV are larger than $\lambda=0.618 + 0.043$ for all pseudo-rapidity windows. 
%--------------------------------------------------------------------

\begin{figure}[h!]
  \begin{center}
      \includegraphics[width=6.8cm,clip]{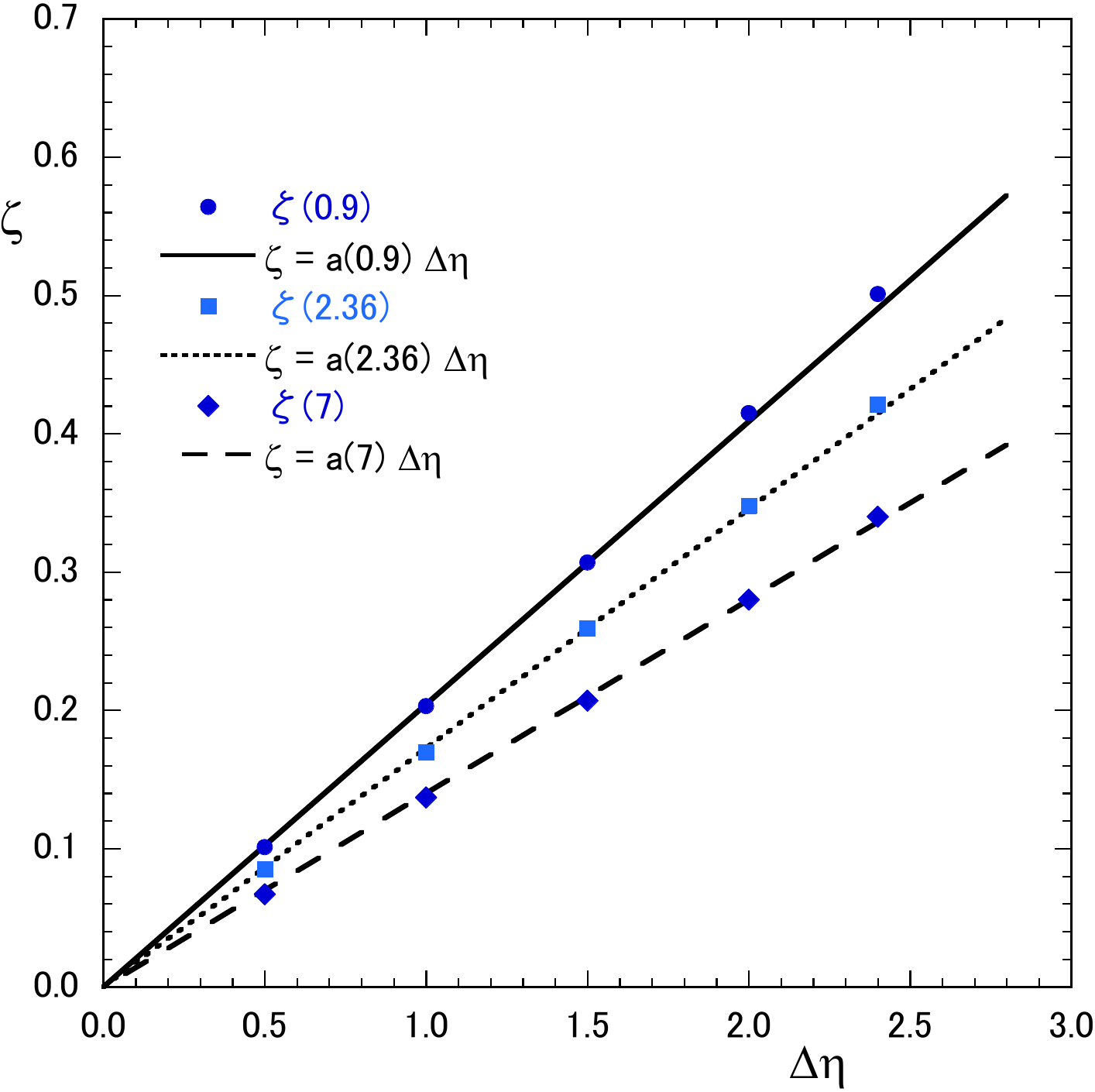}
  \end{center}
  \caption{$\Delta \eta$ dependence of $\zeta$ estimated from the analysis of CMS data at $\sqrt{s} = 0.9, 2.36$ and $7$ TeV. }
  \label{fig.cms_GL2} 
\end{figure}

The pseudo-rapidity window $\Delta \eta$ dependence of estimated values of probability $\zeta$ shown in Fig.\ref{tab.cms_GL} are fitted by a straight line, $\zeta = a\, \Delta \eta$, at each $\sqrt{s}$.
Results are shown in  Fig.\ref{fig.cms_GL2} and 
estimated values of slope parameter $a$ are listed in Table \ref{tab.cms_slope_para}. 

 \begin{table}[htb]
  \caption{Slope parameters estimated from the analysis of charged multiplicity 
 distributions at $\sqrt{s} = 0.9, 2.36$ and $7$ TeV by the CMS Collaboration. }
  \label{tab.cms_slope_para}
  {\small
 \begin{center}
  \begin{tabular}{cccccc} \hline \hline
  $\sqrt{s}$ (TeV) &  0.9 & 2.36 &  7 \\ 
   \hline
     $a$  & $0.204 \pm 0.001$ & $0.173 \pm 0.001$  & $0.140 \pm 0.001$ \\ 
     $\chi^2_{\rm min}/{\rm n.d.f.}$  & $26.0/(5-1)$ & $7.0/(5-1)$  & $43.0/(5-1)$ \\   
  \hline \hline
   \end{tabular}  
 \end{center}
 }
 \end{table}
%

\begin{figure}
  \centering
    \includegraphics[width=6.8cm,clip]{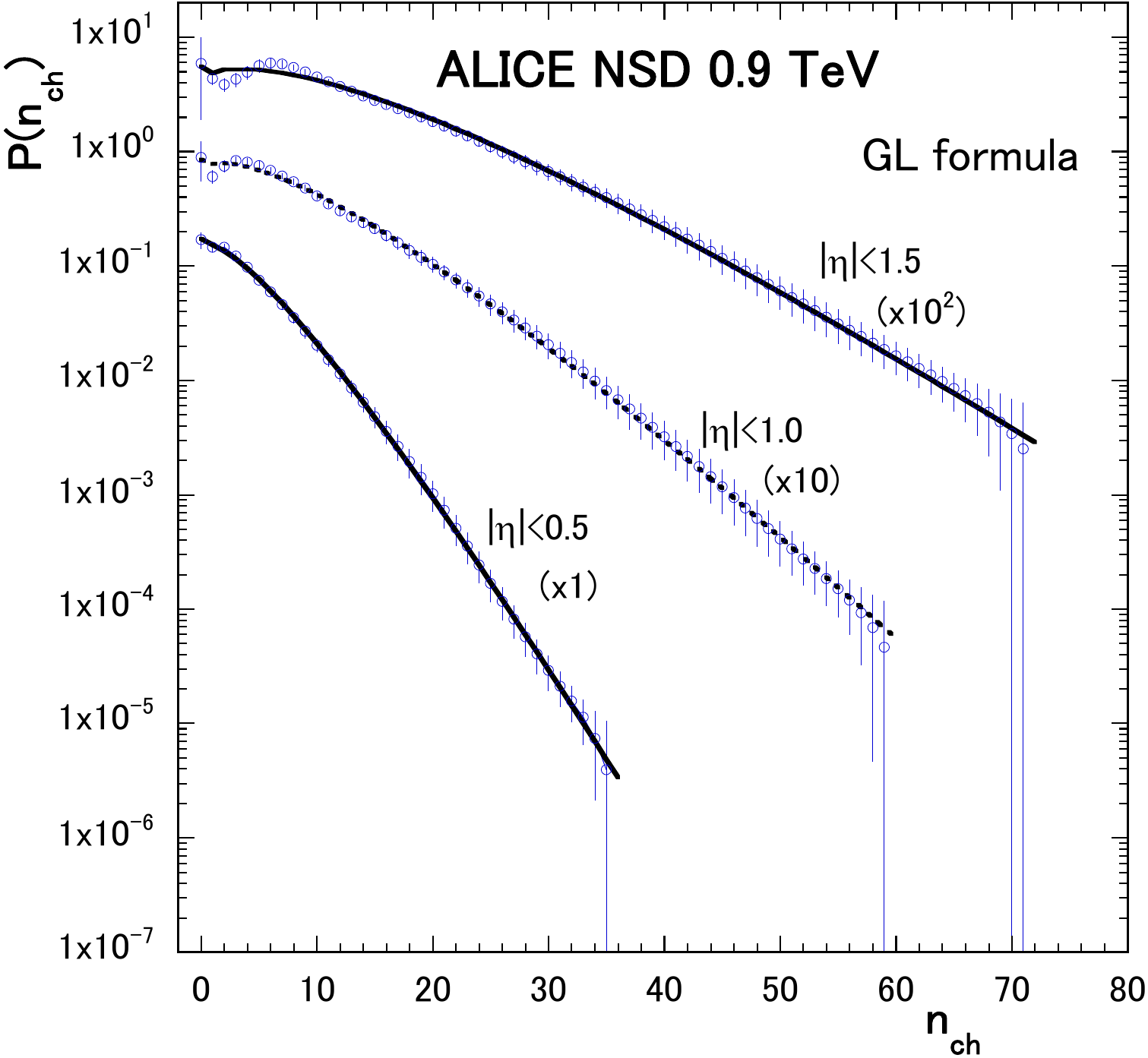}
    \hspace{3mm}
%  \begin{minipage}{0.45\textwidth}
%   \begin{center}
    \includegraphics[width=6.8cm,clip]{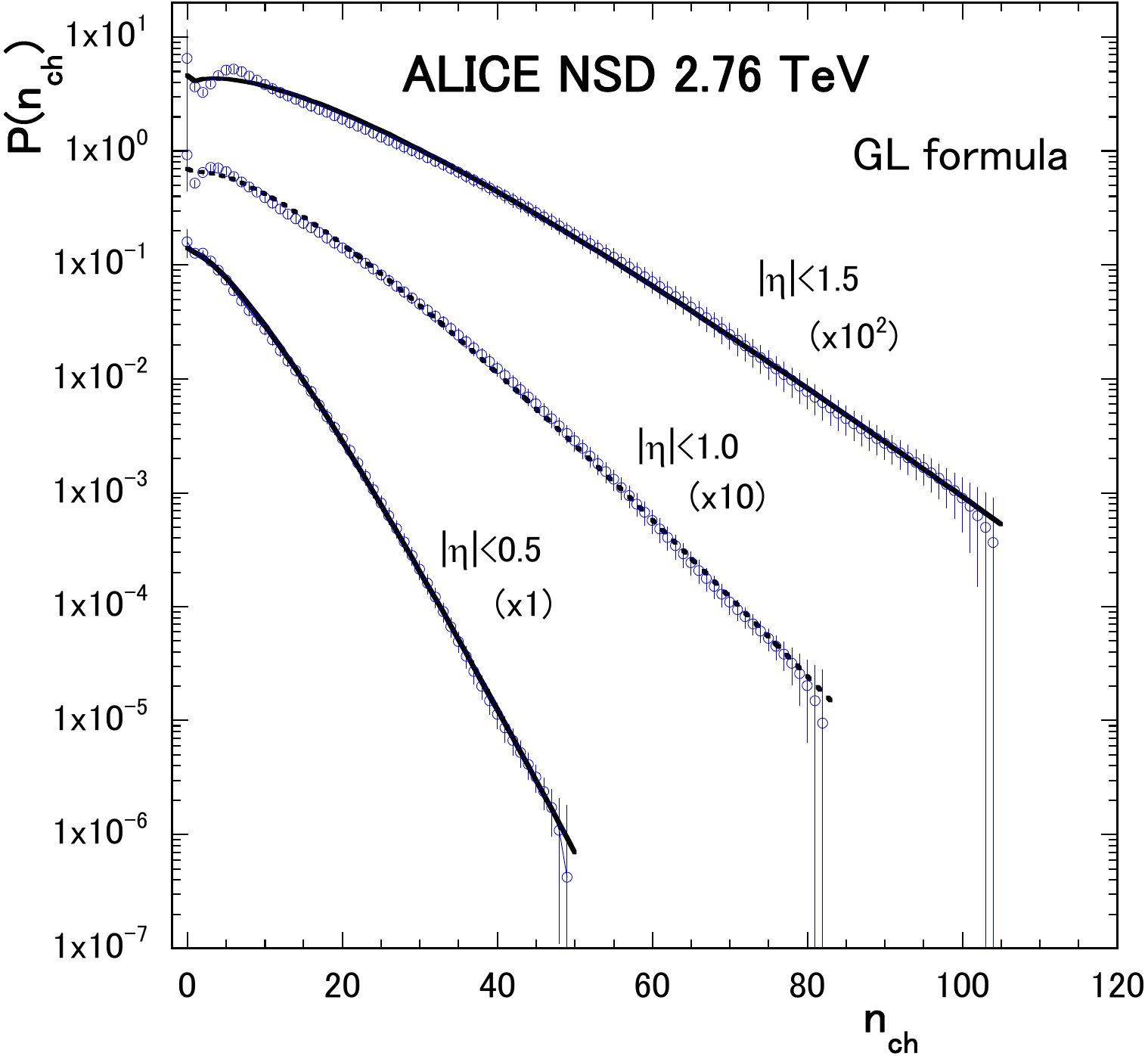}
%   \end{center}
%  \end{minipage} 
  \caption{Charged multiplicity distributions at $\sqrt{s}=0.9$ and $2.76$ TeV 
in the ALICE Collaboration compared to theoretical curves (solid or dotted lines) calculated with Eqs.(\ref{eq.tl06}) and (\ref{eq.anal04}). }
  \label{fig.alice_GL1} 
\end{figure}

\begin{figure}[htb]
 \centering
   \includegraphics[width=6.8cm,clip]{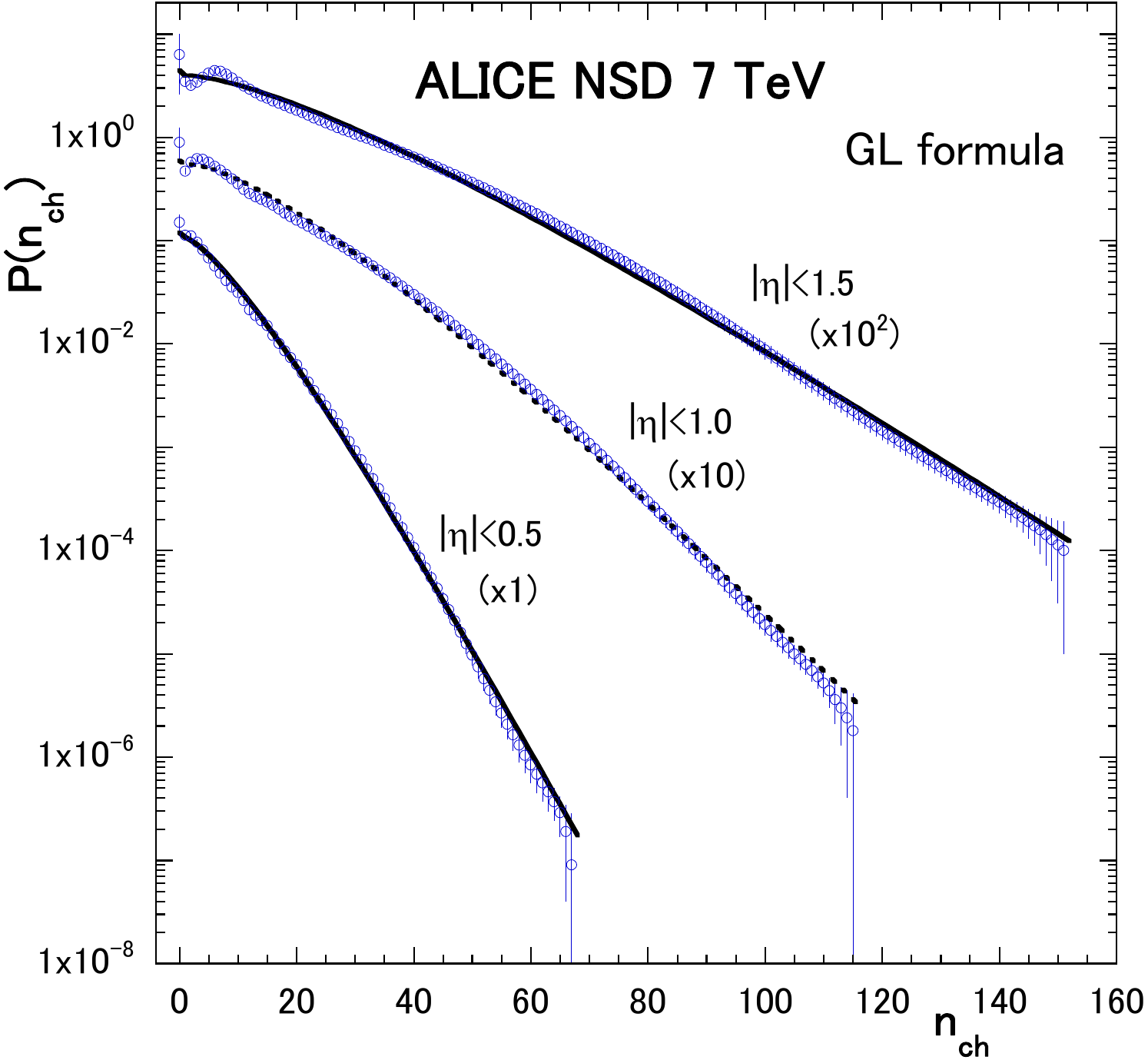}
   \hspace{3mm}
   \includegraphics[width=6.8cm,clip]{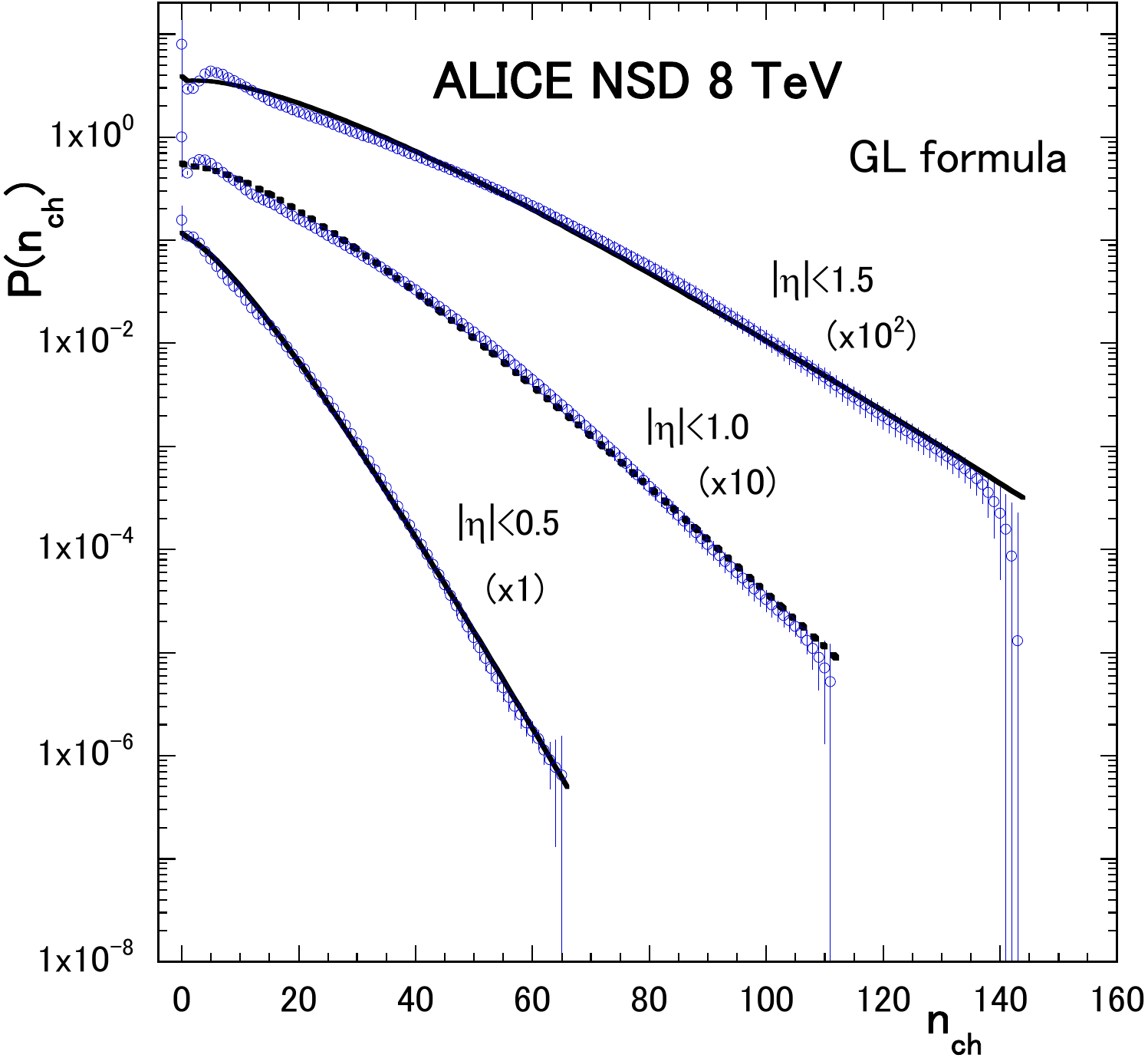}
  \caption{Charged multiplicity distributions at $\sqrt{s}=7$ and  $8$ TeV compared to theoretical curves (solid or dotted lines) calculated with Eqs.(\ref{eq.tl06}) 
and (\ref{eq.anal04}). }
  \label{fig.alice_GL2} 
\end{figure}

%------ ALICE -----
Results on the analysis of the charged multiplicity distributions at $\sqrt{s}=0.9$, 
$2.76$, $7$ and $8$ TeV by the ALICE Collaboration by Eq.(\ref{eq.tl06}) 
and the GL formula, Eq.(\ref{eq.anal04}),  
are shown in Fig.\ref{fig.alice_GL1} and \ref{fig.alice_GL2}. Parameters estimated in the analysis are listed in Table \ref{tab.alice_GL}.

%----- ALICE Fig. -----

At $\sqrt{s}=0.9$ and $2.76$ TeV, values of $\chi^2_{\rm min}/{\rm n.d.f.}$ are less than 1 for three pseudo-rapidity windows,  $|\eta|<0.5$,  $|\eta|<1.0$ and $|\eta|<1.5$.  Calculated results describe the data at
 $0.9$ and $2.76$ TeV by the ALICE Collaboration very well. 
 
 At $\sqrt{s}=7$ TeV, values of $\chi^2_{\rm min}/{\rm n.d.f.}$ are less than 2 except for 2.03 for $|\eta|<1.0$. 
 At $\sqrt{s}=8$ TeV, values of $\chi^2_{\rm min}/{\rm n.d.f.}$ are less than 2. 
 
 %------------
In the analyses of the data by CMS and ALICE Collaborations, results at $\sqrt{s}=7$ and 8 TeV are not better than those at $\sqrt{s}=0.9$ TeV to 2.76 TeV. 
In addition, though, value of $P_{\rm ob}(0)$ satisfies the condition, 
$P_{\rm ob}(0)>P_{\rm ob}(1)$ for each calculation, each peak of measured multiplicity distribution $P_{\rm ch}(n)$ for $|\eta|<\Delta\eta$ with $\Delta \eta \ge 1.0$,   
located around $4<n<8$, cannot be reproduced by the single GL formula. 
In the next subsection, we would analyze the measured multiplicity distributions for $|\eta|<2.4$ at $\sqrt{s}=7$ TeV by the CMS Collaboration and that for $|\eta|<1.5$ at $\sqrt{s}=8$ TeV by the ALICE Collaboration using double GL formulae.

%---------- ALICE Fig -----------
%
%

%---------- ALICE tab -----------
%\newpage
%
 \begin{table}[h!]
  \caption{Parameters estimated from the analysis of charged multiplicity distributions at $\sqrt{s}=0.9, 2.76, 7$ and $\sqrt{s}=8$ TeV by the ALICE 
 Collaboration with  Eqs.(\ref{eq.tl06}) and (\ref{eq.anal04}). }
  \label{tab.alice_GL}
 \begin{center}
 {\small
  \begin{tabular}{ccccccc} \hline \hline
  $\sqrt{s}$ (TeV) &  $\Delta \eta$ &  $p_{\rm in}$  & $\zeta$ 
                         & $\chi_{\rm min}^2/{\rm n.d.f}$ &  $p_{\rm in}(2-p_{\rm in})$ 
                         & $\langle n\rangle_{\rm ob}=2\zeta\langle n\rangle$ \\  \hline
     $0.9$ & $0.5$ & $0.416\pm 0.008$  & $0.106\pm 0.001$ & $4.1/(36-2)$  
                                                    & $0.659\pm 0.005$ & $3.80\pm 0.03$ \\ 
             & $1.0$ & $0.382\pm 0.008$  & $0.217\pm 0.002$ & $15.1/(60-2)$ 
                                                    & $0.618\pm 0.005$ & $7.77\pm 0.06$ \\     
             & $1.5$ & $0.362\pm 0.009$  & $0.328\pm 0.003$ & $38.6/(72-2)$ 
                                                    & $0.593\pm 0.006$ & $11.7\pm 0.1$ \\ 
  \hline 
   $2.76$ & $0.5$ & $0463\pm 0.007$  & $0.0817\pm 0.0005$ & $ 11.7/(50-2)$  
                                                   & $0.693\pm 0.004$ & $4.83\pm 0.03$ \\
             & $1.0$ & $0.415\pm 0.008$ & $0.168\pm 0.001$ & $47.6/(83-2)$ 
                                                   & $0.658\pm 0.006$ & $9.78\pm 0.07$ \\     
             & $1.5$ & $0.396\pm 0.008$ & $0.253\pm 0.002$ & $81.2/(105-2)$ 
                                                   & $0.635\pm 0.006$ & $14.7\pm 0.1$ \\
  \hline   
    $7$    & $0.5$ & $0.483\pm 0.009$ & $0.068\pm 0.001$ & $ 75.2/(68-2)$  
                                                   & $0.733\pm 0.005$ & $6.04\pm 0.05$ \\ 
             & $1.0$ & $0.452\pm 0.009$ & $0.137\pm 0.001$ & $231.5/(116-2)$ 
                                                   & $0.700\pm 0.005$ & $12.2\pm 0.1$ \\     
             & $1.5$ & $0.486\pm 0.008$ & $0.200\pm 0.001$ & $248.7/(152-2)$ 
                                                   & $0.736\pm 0.004$ & $17.8\pm 0.1$ \\ 
  \hline 
     $8$   & $0.5$ & $0.495\pm 0.011$ & $0.066\pm 0.001$ & $ 44.7/(66-2)$ 
                                                   & $0.745\pm 0.006$ & $6.24\pm 0.05$ \\ 
             & $1.0$ & $0.447\pm 0.009$ & $0.135\pm 0.001$ & $137.5/(112-2)$ 
                                                   & $0.694\pm 0.005$ & $12.8\pm 0.1$ \\     
             & $1.5$ & $0.447\pm 0.008$ & $0.201\pm 0.002$ & $209.0/(144-2)$ 
                                                   & $0.694\pm 0.004$ & $19.0\pm 0.1$ \\ 
 \hline  \hline
  \end{tabular}
 } 
 \end{center}
% }
 \end{table}
% 

% -----------------
\subsection{Analysis of charged multiplicity distributions with double GL formulae}
%-------------------------------------------------------------------

In the invariant energy $\sqrt{s}$ region above several hundred GeV, it is assumed that mainly two production processes occur exclusively each other. Process 1 (soft process) occurs with a probability $\alpha$ and the multiplicity distribution of negative particles is given $P_1(n, \langle n_1 \rangle)$, process 2 (semi-hard process)  occurs with a probability ($1-\alpha$) and the multiplicity distribution of 
 negative particles is given $P_2(n, \langle n_2 \rangle)$
In the full phase space, combined multiplicity distribution $P(n, \langle n \rangle)$ 
can be given by the following equation,
 \begin{eqnarray}
   P(n, \langle n\rangle) = \alpha P_1(n, \langle n_1\rangle) 
                       + (1-\alpha)  P_2(n, \langle n_2\rangle).  \label{eq.dgl01}
 \end{eqnarray}
From Eq.(\ref{eq.dgl01}), we obtain
 \begin{eqnarray}
    \langle n\rangle = \alpha \langle n_1\rangle + (1-\alpha) \langle n_2\rangle.
     \label{eq.dgl02}
 \end{eqnarray}

In our approach, the observed multiplicity distribution  $P_{\rm ob}(n)$ in a pseudo-rapidity window is given by 
 \begin{eqnarray}
    &&   P_{\rm ob}(n)  = \alpha GL_1(n) + (1-\alpha) GL_2(n),  \nonumber \\
    &&  GL_i(n) = \sum_{j=0}^{[n/2]} {}_{n-j} C_{j}\,
          \frac{ ({\zeta_i}^2)^{j} [ 2\zeta_i(1-\zeta_i) ]^{n-2j} }
                  { [ \zeta_1(2-\zeta_1) ]^{n-j} } 
                               P(n-j, \langle n_{i\zeta_i} \rangle), \quad i=1,2,    
          \label{eq.ap_dgl03}          
 \end{eqnarray}
where $\langle n_{i\zeta_i} \rangle = \zeta_i(2-\zeta_i) \langle n_i \rangle$.
We assume that $\langle n_1\rangle > \langle n_2 \rangle$ and that each multiplicity distribution $P_i(n, \langle n_i\rangle)$ is given by the Glauber-Lachs (GL) formula, 
 \begin{eqnarray}
     P(n,\langle n_i \rangle) 
        = \frac{(p_i \langle n_i \rangle)^n}{(1+p_i\langle n_i\rangle)^{n+1}} 
           \exp\Bigl[-\frac{(1-p_i) \langle n \rangle }{1+p_i \langle n_i \rangle} \Bigr]
           {L_n}\Bigl( -\frac{(1-p_i) }{ p_i(1+p_i \langle n_i \rangle) }  \Bigr).  
            \label{eq.ap_dgl04} 
 \end{eqnarray}
We parametrize as $\langle n_i \rangle = r_i \langle n \rangle$ ($i=1,2$). 
% 
% \begin{eqnarray*}
%   \langle n_i \rangle = r_i \langle n \rangle, \quad i=1,2.
% \end{eqnarray*}
% 
Then, we obtain
 \begin{eqnarray}
      1= \alpha r_1 + (1-\alpha) r_2, \quad  r_1>r_2>0.    \label{eq.ap_dgl05}     
 \end{eqnarray}
In our parametrization, $\langle n \rangle$ is given from Eq.(\ref{eq.anal01}) 
or Table \ref{tab.avmlt}, and $r_2$ is determined from Eq.(\ref{eq.ap_dgl05}). 
Therefore, 6 parameters $\alpha$, $r_1$, $p_1$, $\zeta_1$ $p_2$ and $\zeta_2$, 
are contained in Eq.(\ref{eq.ap_dgl03}).

\begin{figure}[htb] 
  \centering
    \includegraphics[width=6.8cm,clip]{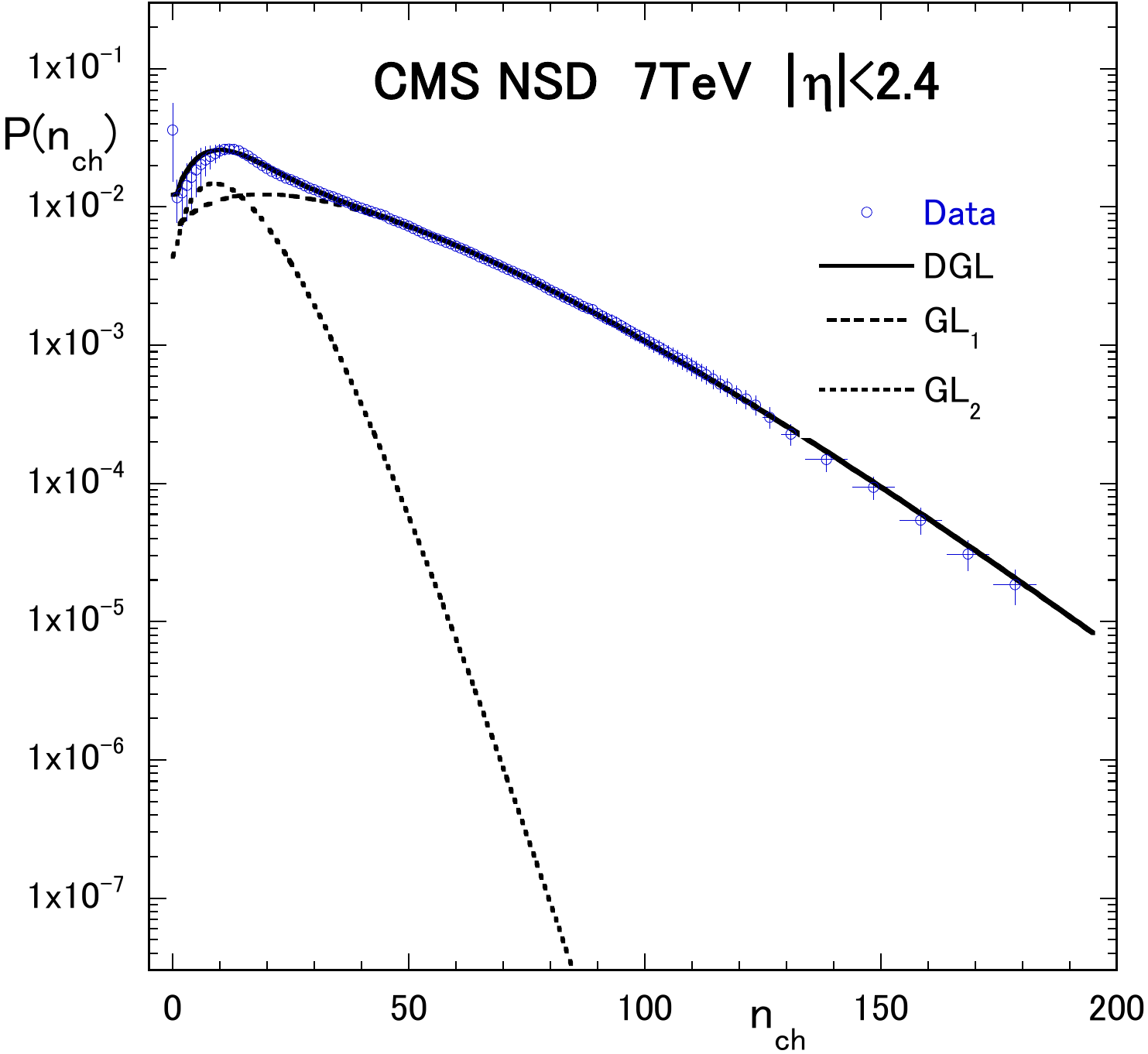}
    \hspace{3mm}
    \includegraphics[width=6.8cm,clip]{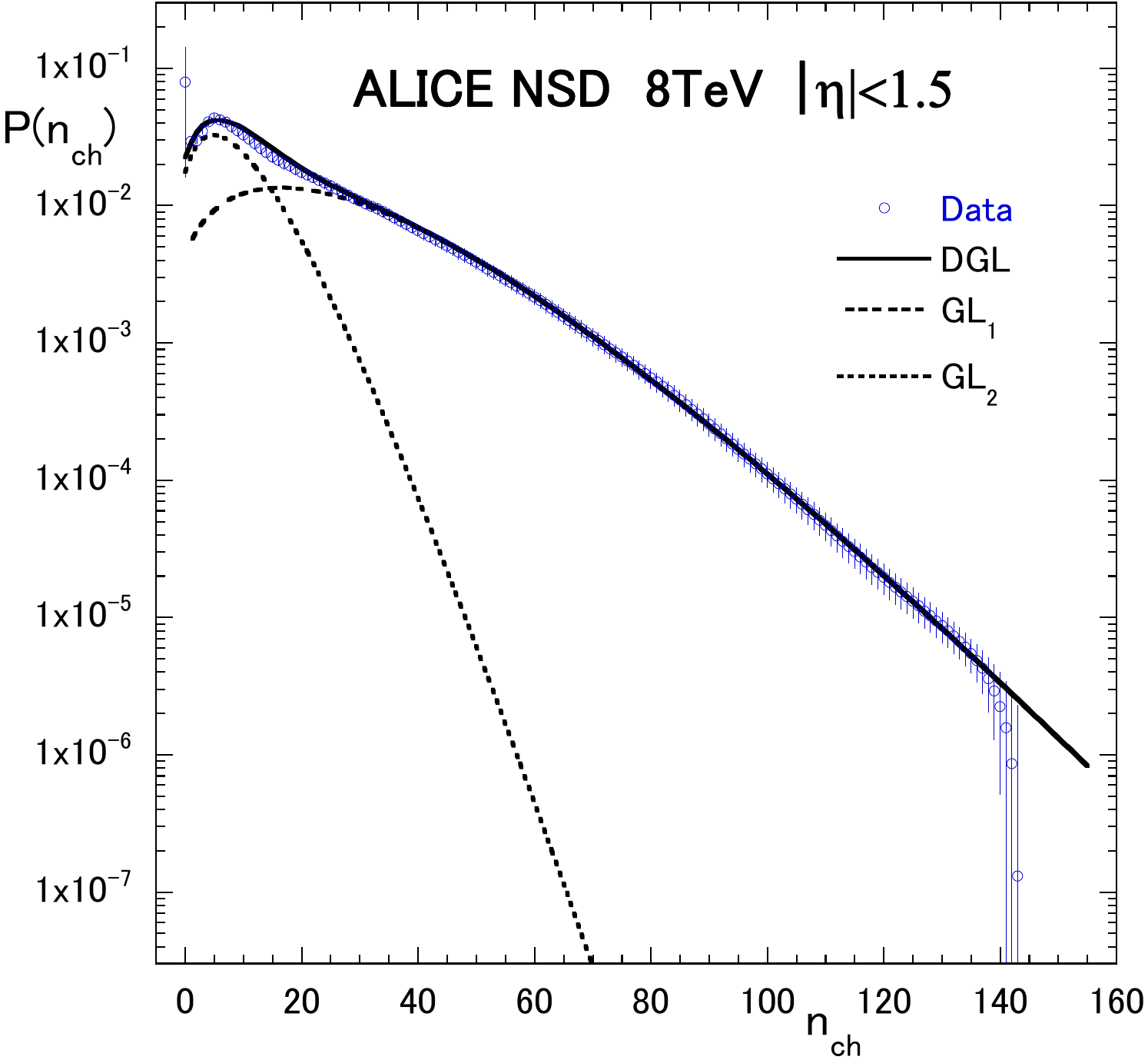}
  \caption{Charged multiplicity distributions at $\sqrt{s}=7$ TeV by the CMS Collaboration and  that at $\sqrt{s}=8$ TeV by the ALICE Collaboration compared to  theoretical curves (solid or dotted lines) calculated with parameters shown in Table \ref{tab.anal_DGL}. }
  \label{fig.ana_DGL} 
\end{figure}

Results on the analyses of measured charged multiplicity distribution 
for $|\eta|<2.4$ at 7 TeV by the CMS Collaboration and that for $|\eta|<1.5$ 
at 8 TeV by the ALICE Collaboration with the double GL formulae 
are shown in Fig.\ref{fig.ana_DGL}. 
Parameters estimated form the analyses are listed in Table \ref{tab.anal_DGL}. 

 \begin{table}[htb]
  \caption{Parameters estimated from the analysis of charged multiplicity distributions at $\sqrt{s}=7$ TeV by the CMS Collaboration and at $8$ TeV by the ALICE Collaboration by the double GL formulae. }
  \label{tab.anal_DGL}
 \begin{center}
  \begin{tabular}{cccccc} \hline \hline
  $\sqrt{s}$(TeV) & $\Delta\eta$ & $\alpha$ &  $p_1$  & $\zeta_1$ & $r_1$  \\
 \hline
   $7$ & 2.4 & $0.717\pm 0.022$& $0.290\pm 0.012$& $0.330\pm 0.017$& $1.28\pm 0.01$ \\ 
% \hline
   $8$ & 1.5 & $0.540\pm 0.045$& $0.228\pm 0.017$& $0.204\pm 0.168$& $1.41\pm 0.12$ \\ 
 \hline \hline
  \end{tabular}
 \end{center}
 \end{table}
 \begin{table}[htb]
 \begin{center}
  Table \ref{tab.anal_DGL}.  \textit{(Continued)}.  \\
  \begin{tabular}{cccc} \hline \hline
        $p_2$        &   $\zeta_2$ &  $r_2$ & $\chi_{\rm min}^2/{\rm n.d.f}$  \\   \hline
   $0.158\pm 0.028$ & $0.509\pm 0.279$& $0.291\pm 0.002$& $16.2/(127-6)$ \\ 
% \hline
   $0.223\pm 0.036$& $0.179\pm 0.471$& $0.519\pm 0.018$& $50.8/(144-6)$ \\ 
 \hline \hline
  \end{tabular}
 \end{center}
 \end{table}
%

%%-------
\section{Concluding remarks}
%%------------------------

Multiplicity distribution in the pseudo-rapidity window, which satisfies the charge conservation in the full phase space, is formulated. By the use of the GL formula for 
the multiplicity distribution of negative charged particles in the full phase space, 
we analyze the charged multiplicity distributions in pseudo-rapidity windows in non-single diffractive (NSD) events reported by CMS and ALICE Collaborations.

R1) The probability $\zeta$ that each particle enter into the given pseudo-rapidity window  $|\eta|<\Delta\eta$ is approximately expressed by $\zeta=a \Delta\eta$ 
 with parameter $a$, which depends on the invariant energy $\sqrt{s}$.

R2)  In our analysis, relation, $P_{\rm ob}(0) > P_{\rm ob}(1) $, holds, which is similar to the experimental data.  In \cite{Rybczynski2018}, relation $P_{\rm ob}(0) > P_{\rm ob}(1) $ and peak around $4<n<8$ are well reproduced by the use of a compound  distribution.   For the relation $P_{\rm ob}(0) > P_{\rm ob}(1) $, see also \cite{Pumplin1994}.

R3) In the measured charged multiplicity distributions for $\Delta \eta > 1.5$,  a peak appears around $4 < n <8$ in each distribution. We cannot reproduce the peak from our calculation with the single GL formula. 

R4) We can reproduce global behavior of measured multiplicity distributions for $\Delta\eta=2.4$ at $\sqrt{s}=7$ TeV by the CMS Collaboration, and   for $\Delta\eta=1.5$ at $\sqrt{s}=8$ TeV by the ALICE Collaboration with the double GL formulae.

R5) For example, if two jet-like structure appears and charge conservation is satisfied in each jet-like structure in the soft or semi-hard process~\cite{Capella1994}, it would be appropriate to use the GGL formula with k=2.

R6) We obtain the relation, $P_{\rm ob}(n) \simeq P(n, 2\zeta \langle n \rangle)$,  
from Eqs.(\ref{eq.tl05}) and (\ref{eq.tl06}), if $\zeta^2$ is much smaller than $\zeta$.

%--------
\appendix
%---------------------------------------------------------
%----------------
 \section{Multiplicity distribution in a pseudo-rapidity window}
 \label{apdx.mdcc}
%---------------

In the full phase space, the measured multiplicity distribution should satisfy the charge conservation. For simplicity, we assume that the charged particles are produced in pairs of a positive charged particle and a negative charged particle. Let $P(n)$, $n=0,1,\ldots$ be a multiplicity distribution of negative charged particles, and $P_{\rm ch}(2n)$ be that of charged particles in the full phase space. We assume that a relation, 
 \begin{eqnarray}
    P(n)=P_{\rm ch}(2n) \label{eq.ap_qoa01}
 \end{eqnarray}
holds. The probability generating function (GF) $\Pi(z)$ for $P(n)$, and that $\Pi_{\rm ch}(z)$ for $P_{\rm ch}(2n)$ are respectively written as,  
 \begin{eqnarray}
    \Pi(z) = \sum_{n=0}^\infty P(n) z^n, \quad 
    \Pi_{\rm ch}(z) = \sum_{n=0}^\infty P_{\rm ch}(2n) z^{2n}.      
    \label{eq.ap_qoa02}         
 \end{eqnarray}
From Eq.(\ref{eq.ap_qoa02}), we have following relations,
 \begin{eqnarray*}
   && P(n)=\frac{1}{n!}\Pi^{(n)}(z)|_{z=0}, \\ 
   && P_{\rm ch}(2n-1) = \frac{1}{(2n-1)!} \Pi_{\rm ch}^{(2n-1)}(z)|_{z=0} = 0,  \quad
        P_{\rm ch}(2n) = \frac{1}{(2n)!} \Pi_{\rm ch}^{(2n)}(z)|_{z=0}.
 \end{eqnarray*}
From Eqs.(\ref{eq.ap_qoa01}) and (\ref{eq.ap_qoa02}, the following relation is satisfied:
 \begin{eqnarray}
    \Pi_{\rm ch}(z) = \Pi(z^2).        \label{eq.ap_qoa03}           
 \end{eqnarray}

It is assumed that a probability that each particle produced in the full phase space enters into a pseudo-rapidity window is $\zeta$ ( $0\le \zeta \le 1$), and that each particle does not enter into the window is $1-\zeta$.  
When more than $n$ pairs of positive and negative charged particles are produced 
in the full phase space, and $m$ ($2n\ge m\ge 0$) charged particles enter into the 
 pseudo-rapidity window, the probability distribution that $m$ charged particles are detected, $P_{\rm ob}(m)$, is written as
 \begin{eqnarray}
    P_{\rm ob}(m) = \sum_{2n\ge m}^\infty {}_{2n}C_m 
                  \zeta^m(1-\zeta)^{2n-m} P_{\rm ch}(2n).        \label{eq.ap_qoa04}
 \end{eqnarray}
The GF for $P_{\rm ob}(m)$ is defined by
 \begin{eqnarray}
    \Pi_{\rm ob}(z) = \sum_{m=0}^\infty P_{\rm ob}(m) z^m.  \label{eq.ap_qoa05}
 \end{eqnarray}
Substituting Eq.(\ref{eq.ap_qoa04}) into Eq.(\ref{eq.ap_qoa05}), and using the definition of $\Pi_{\rm ch}(z)$, Eq.(\ref{eq.ap_qoa02}), we obtain 
 \begin{eqnarray}
   \Pi_{\rm ob}(z) =  \Pi\bigl( (\zeta z + 1-\zeta)^2 \bigr).      \label{eq.ap_qoa06} 
 \end{eqnarray}

Putting
 \begin{eqnarray}
       y = \frac{ (\zeta z)^2 + 2\zeta(1-\zeta)z }{\zeta(2-\zeta)},     
\label{eq.ap_qoa07} 
 \end{eqnarray}
and using the relation, $ (\zeta z + 1-\zeta)^2=\zeta(2-\zeta)y +(1-\zeta)^2$, we can rewrite Eq.(\ref{eq.ap_qoa06}) as
 \begin{eqnarray*}
    && \Pi_{\rm ob}(z) =  \Pi \bigl( \zeta(2-\zeta)y + (1-\zeta)^2 \bigr)
             = \sum_{n=0}^{\infty} \bigl[ \zeta(2-\zeta)y + (1-\zeta)^2  \bigr]^n P(n) \\
    && \hspace{12mm} = \sum_{j=0}^\infty y^j \sum_{n=j}^{\infty} 
          {}_nC_j\,  \bigl[ \zeta(2-\zeta) \bigr]^j \bigl[ (1-\zeta)^2 \bigr]^{n-j} P(n).
 \end{eqnarray*}

We define the multiplicity distribution $P_\zeta(j)$ as 
 \begin{eqnarray}
      P_\zeta (j) \equiv \sum_{n=j}^{\infty} {}_nC_j\,  
           \bigl[ \zeta(2-\zeta) \bigr]^j \bigl[ (1-\zeta)^2 \bigr]^{n-j} P(n), 
            \quad j=0,1,2,\cdots.                 \label{eq.ap_qoa08} 
 \end{eqnarray}
Equation (\ref{eq.ap_qoa08}) denotes the probability that when $n$ pairs  ($n\ge j$) of charged particles are produced, 
$(n-j)$ pairs are outside the pseudo-rapidity window, 
and at least one particle enters into the window from any $j$ pairs of negative 
and positive charged  particles.   
 The GF $\Pi_\zeta(y)$ is defined as
 \begin{eqnarray}
    &&  \Pi_\zeta(y) = \sum_{j=0}^\infty y^j P_\zeta(j). \label{eq.ap_qoa09} 
 \end{eqnarray}

In the following, the multiplicity distribution $P(n)$ is written as $P(n,\langle n \rangle)$, where $\langle n \rangle$ is the average multiplicity of negative charged particles in the full phase space. It's GF $\Pi(x)$ is also written as $\Pi(x,\langle n\rangle)$:
 \begin{eqnarray*}
    &&  \Pi(x, \langle n \rangle) = \sum_{j=0}^\infty P(j,\langle n \rangle)x^j.
 \end{eqnarray*}

Then, we obtain two relations among three GF's: 
 \begin{eqnarray}
    &&  \Pi_{\rm ob}(z) = \Pi_\zeta(y), \quad 
      y = \frac{ (\zeta z)^2 + 2\zeta(1-\zeta)z}{\zeta(2-\zeta)}, \label{eq.ap_qoa10} \\
    &&  \Pi_\zeta(y) = \Pi(x,\langle n \rangle), \quad x=\zeta(2-\zeta)(y-1)+1. \label{eq.ap_qoa11} 
 \end{eqnarray}

It should be noted that $\Pi_{\rm ob}(z)$ is the GF for  
$P_{\rm ob}(n)$,  $\Pi_\zeta(y)$ is that for $P_\zeta(n)$, 
and $\Pi(x, \langle n \rangle)$ is for $P(n, \langle n \rangle)$.

%%---------------
\subsection{Relation between $P_{\rm ob}(n)$ and $P_\zeta(n)$}
%%------------------------------------------------------

We define  $f_{\rm ob}^{(n)}(z)$ and $f_{\zeta}^{(n)}(y)$  respectively as
 \begin{eqnarray}
     f_{\rm ob}^{(n)}(z) 
                  = \frac{1}{n!} \frac{\partial^n}{\partial z^n} \Pi_{\rm ob}^{(n)}(z),   \quad  
     f_{\zeta}^{(n)}(y) = \frac{1}{n!}\frac{\partial^n}{\partial y^n} \Pi_\zeta(y),  
          \label{eq.ap_ob01}  
 \end{eqnarray}
where
 \begin{eqnarray}
    &&  y = rz^2 + qz, \quad r = \zeta^2/[ \zeta(2-\zeta) ], 
          \quad q= 2\zeta(1-\zeta)/[ \zeta(2-\zeta) ].  \label{eq.ap_ob02}
 \end{eqnarray}
From Eq.(\ref{eq.ap_qoa10}), we can show that the following equation is satisfied:
 \begin{eqnarray}
    f_{\rm ob}^{(n)}(z) = \sum_{j=0}^{[n/2]} {}_{n-j}C_j r^{j} 
                   (\partial y/\partial z)^{(n-2j)} f_\zeta^{(n-j)}(y).     \label{eq.ap_ob05}
 \end{eqnarray}
From the definition of the GF, we obtain that 
$P_{\rm ob}(n) = f_{\rm ob}^{(n)}(z)|_{z=0}$ and  
$P_\zeta(n) = f_\zeta^{(n)}(y)|_{y=0}$.

If $z=0$, then $y=0$ from Eq.(\ref{eq.ap_ob02}). Therefore, we obtain from Eq.(\ref{eq.ap_ob05}):
 \begin{eqnarray}
       P_{\rm ob}(n)  = \sum_{j=0}^{[n/2]} {}_{n-j} C_{j}\,
           \frac{ (\zeta^2)^{j} [ 2\zeta(1-\zeta) ]^{n-2j} }{ [ \zeta(2-\zeta) ]^{n-j} } 
                               P_\zeta(n-j).    \label{eq.ap_ob06}          
 \end{eqnarray}
% 

%%-------------------
\subsection{Relation between $P_\zeta(n)$ and $P(n, \langle n \rangle)$, or $\Pi_\zeta(y)$ 
and $\Pi(x, \langle n \rangle)$}
%%-------------------

If variable $x$ is contained in the form of $\langle n\rangle(x-1)$ in $\Pi(x,\langle n\rangle)$,   $\Pi_\zeta(y)$ in Eq.(\ref{eq.ap_qoa11}) is written as
 \begin{eqnarray}
    &&  \Pi_\zeta(y) = \Pi(y,\langle n_\zeta \rangle), \quad 
          \langle n_\zeta \rangle =\zeta(2-\zeta)\langle n\rangle. \label{eq.ap_ob11} 
 \end{eqnarray}

For example, let the multiplicity distribution of negative charged particles be given by the Generalized Glauber-Lachs (GGL) formula with $0<p<1$,
 \begin{eqnarray}
     P(n,\langle n \rangle) 
        = \frac{(p \langle n \rangle/k)^n}{(1+p\langle n\rangle/k)^{n+k}} 
           \exp\Bigl[-\frac{(1-p) \langle n \rangle }{1+p \langle n \rangle/k} \Bigr]
           {L_n}^{(k-1)}\Bigl( -\frac{(1-p)k }{ p(1+p \langle n \rangle/k) }  \Bigr).  
            \label{eq.ap_ob12} 
 \end{eqnarray}
Its generating function is given by
 \begin{eqnarray}
     \Pi(x,\langle n \rangle) = \sum_{n=0}^\infty P(n,\langle n \rangle) x^n 
       =  \Bigl(1-\frac{p\langle n\rangle}{k} (x-1) \Bigr)^{-k} 
                \exp\Bigl[\frac{(1-p) \langle n \rangle(x-1) }
                                       {1-\frac{p \langle n \rangle}{k} (x-1) } \Bigr].  
          \label{eq.ap_ob13} 
 \end{eqnarray}
When $k=1$ with $p=p_{\rm in}$, the GGL formula, Eq.(\ref{eq.ap_ob12}), 
reduces to the GL formula, Eq.(\ref{eq.int05}). 
Then, the generating function $\Pi_\zeta(y)$ for $P_\zeta(n)$ is given from 
Eq.(\ref{eq.ap_ob11}) as
 \begin{eqnarray}
      \Pi_\zeta(y) = \Pi(y,\langle n_\zeta \rangle) 
         =  \Bigl(1- \frac{p\langle n_\zeta\rangle}{k}(y-1) \Bigr)^{-k} 
             \exp\Bigl[\frac{(1-p) \langle n_\zeta \rangle(y-1) }
                             {1- \frac{p \langle n_\zeta \rangle }{k} (y-1) } \Bigr].  
       \label{eq.ap_ob14} 
 \end{eqnarray}

The multiplicity distribution $P_\zeta(n)$ is given from 
$\Pi(y, \langle n_\zeta\rangle)$, and is equal to $P(y,\langle n_\zeta \rangle)$:
 \begin{eqnarray*}
   P_\zeta(n) 
   = \frac{1}{n!} \frac{\partial^n \Pi(y, \langle n_\zeta\rangle) }{\partial y^n}\Big|_{y=0} 
             = P(n,\langle n_\zeta \rangle).
 \end{eqnarray*}
Therefore,  $P_\zeta(n)$ is given from Eq.(\ref{eq.ap_ob12}), if $\langle n\rangle$ 
is replaced by $\langle n_\zeta \rangle$.

In the limit of $p=0$, the generating function $\Pi(x,\langle n \rangle)$ in Eq.(\ref{eq.ap_ob13}) reduces to that of the PSND,
 \begin{eqnarray*}
     \Pi(x,\langle n \rangle) = e^{ \langle n \rangle(x-1) }.
 \end{eqnarray*}
In the limit of $p=1$, it reduces to the generating function of NBD,
 \begin{eqnarray*}
     \Pi(x,\langle n \rangle) 
       =  \bigl[1- \langle n\rangle (x-1)/k \bigr]^{-k}. 
 \end{eqnarray*}
%

% -----------------------------------------------

The relation among  $P(n,\langle n \rangle)$, $\Pi(x, \langle n \rangle)$ and $\Pi_\zeta(y)$ for the GGL formula is listed  
 in Table \ref{ap_tab1} with other two examples.

 \begin{table}[htb]
  \caption{Relation among $P(n,\langle n\rangle)$, $\Pi(x,\langle n\rangle)$ and $\Pi_\zeta(y)$. }
  \label{ap_tab1}
 \begin{center}
  \begin{tabular}{cccc} \hline \hline
      & $P(n,\langle n\rangle)$ & $\Pi(x,\langle n\rangle)$ & $\Pi_\zeta(y) = \Pi(y, \langle n_\zeta\rangle)$  \\ \hline
 Poisson & $ \displaystyle{ \frac{\langle n \rangle^n}{n!}e^{-\langle n\rangle} } $
                 & $\displaystyle{ e^{\langle n\rangle(x-1)} }$ 
                 & $\displaystyle{ e^{\langle n_\zeta\rangle(y-1)} }$ \\ \hline
 NBD     & $\displaystyle{  \frac{\Gamma(n+k)}{n!\Gamma(k)}\frac{(\langle n \rangle/k)^n}  
                                    {(1+\langle n\rangle/k)^{n+k}}  } $
             & $\displaystyle{ \Bigl( 1-\frac{\langle n\rangle}{k}(x-1) \Bigr)^{-k} } $       
             & $\displaystyle{ \Bigl(1-\frac{\langle n_\zeta\rangle}{k}(y-1) \Bigr)^{-k} } $\\ \hline    
 GGL      & Eq.(\ref{eq.ap_ob12})  & Eq.(\ref{eq.ap_ob13}) & Eq.(\ref{eq.ap_ob14}) \\ \hline    \hline
  \end{tabular}
 \end{center}
 \end{table}
% 

%\newpage
%%---------------
\subsection{Difference between $P_{\rm ob}(n)$ and $P(n)$ in the second order factorial moment}
%%------------------------------------------------------

The $m$th order factorial moments for $P_{\rm ob}(n)$ and $P_\zeta(n)$ are respectively given by
 \begin{eqnarray*}
    F_{m, \rm{ob}} &\equiv & \langle n(n-1) \cdots (n-m+1) \rangle_{\rm ob} 
               = \frac{\partial^m \Pi_{\rm ob}(z)}{\partial z^m}\Big|_{z=1}, \\
    F_{m, \zeta} &\equiv & \langle n(n-1) \cdots (n-m+1) \rangle_{\zeta} 
              = \frac{\partial^m \Pi_{\zeta}(y)}{\partial y^m}\Big|_{y=1}.        
 \end{eqnarray*}
If $z=1$, then $y=1$ from Eq.(\ref{eq.ap_ob02}). Then we obtain from Eq.(\ref{eq.ap_ob05}):
 \begin{eqnarray}
     F_{m, \rm{ob}}  = \sum_{j=0}^{[m/2]} 
       \frac{m!}{j! (m-2j)!} 
           \frac{ (\zeta^2)^{j} [ 2\zeta(1-\zeta) ]^{m-2j} }{ [ \zeta(2-\zeta) ]^{m-j} } 
                        F_{m-j, \zeta}.    \label{eq.ap_ob21}          
 \end{eqnarray}
From Eq.(\ref{eq.ap_ob21}), we obtain, 
 \begin{eqnarray}
    && \langle n \rangle_{\rm{ob}}  = 2 \zeta \langle n \rangle, \label{eq.ap_ob22}  \\
    && \langle n(n-1)\rangle_{\rm{ob}} 
             = [ 2/(2-\zeta) ]^2  \langle n(n-1)\rangle_{\zeta} 
                + 2\zeta\langle n \rangle.      \label{eq.ap_ob23}          
 \end{eqnarray}

In the case of GGL formula, the second order factorial moment for $P_{\rm{ob}}(n)$ is given by
 \begin{eqnarray}
   &&  \langle n(n-1) \rangle_{\rm ob} 
              = [ 1+p(2-p)/k ] [ 2\zeta \langle n \rangle ]^2
               + \zeta [ 2\zeta \langle n \rangle ].  \label{eq.apd101}
 \end{eqnarray}
On the other hand, that for $P(n)$ is given by 
 \begin{eqnarray}
   &&  \langle n(n-1) \rangle 
              = [ 1+p(2-p)/k ] \langle n \rangle^2. \label{eq.apd102}
 \end{eqnarray}
As can be seen from Eqs.(\ref{eq.apd101}) and (\ref{eq.apd102}),  an additional term, 
$\zeta [ 2\zeta  \langle n \rangle ]$, appears on the right hand side of  Eq.(\ref{eq.apd101}), which is caused by the charge conservation in the full phase space.

%----------------------------------------------
\section{Analysis by the convolution of NBD and PSND}
\label{apdx.FW}
%----------------------------------------------

In order to compare the results by Eqs.(\ref{eq.tl06}) and (\ref{eq.anal03}), where charge conservation in the full phase space is taken into account, we also analyze the data by the convolution of NSD and PSND, Eq.(\ref{eq.int01}).
Results at $\sqrt{s}=0.9$ and $2.36$ TeV in the CMS Collaboration are shown respectively in Tables \ref{tab.cms0900FWb} and \ref{tab.cms2360FWb}. 

%---------- cms0900_FW_tab----------
%
 \begin{table}[htb]
  \caption{Parameters estimated from the analysis of charged multiplicity 
 distributions at $\sqrt{s}=0.9$ TeV in the CMS collaboration 
 by Eq.(\ref{eq.int01}), where $\langle n \rangle$ is replaced to 
 $\langle n_{\rm ch}\rangle$. }
  \label{tab.cms0900FWb}
%  {\small
 \begin{center}
  \begin{tabular}{cccccc} \hline \hline
  $\sqrt{s}$ (TeV) &  $k$ & $\Delta \eta$ & $\tilde{p}$ & $\langle n_{\rm ch} \rangle$ 
                                                 & $\chi_{\rm min}^2/{\rm n.d.f.}$ \\ 
 \hline
  $0.9$&$1$& $0.5$ & $0.808\pm 0.029$ & $ 3.64 \pm 0.09$ & $ 70.5/(23-2)$ \\ 
          &     & $1.0$ & $0.689\pm 0.025$ & $ 7.40 \pm 0.19$ & $322.4/(40-2)$ \\     
          &     & $1.5$ & $0.713\pm 0.019$ & $11.3 \pm 0.21$ & $291.3/(52-2)$ \\     
          &     & $2.0$ & $0.716\pm 0.016$ & $15.3 \pm 0.24$ & $290.4/(62-2)$ \\     
          &     & $2.4$ & $0.703\pm 0.015$ & $18.5 \pm 0.30$ & $352.4/(68-2)$ \\ 
 \hline    
  $0.9$&$2$&$0.5$& $1.000\pm 0.039$ & $ 3.90\pm 0.09$ & $ 70.4/(23-2)$  \\ 
          &     &$1.0$& $1.000\pm 0.025$ & $ 7.61\pm 0.12$ & $104.3/(40-2)$ \\     
          &     &$1.5$& $1.000\pm 0.018$ & $11.4 \pm 0.14$ & $ 89.0 /(52-2)$ \\     
          &     &$2.0$& $1.000\pm 0.012$ & $15.3 \pm 0.12$ & $ 60.2/(62-2)$ \\     
          &     &$2.4$& $1.000\pm 0.011$ & $18.3 \pm 0.13$ & $ 66.4/(68-2)$ \\ 
  \hline \hline
   \end{tabular}  
 \end{center}
% }
 \end{table}
%

%---------- cms2360_FW_tab ----------
%
 \begin{table}[htb]
  \caption{Parameters estimated from the analysis of charged multiplicity 
 distributions at $2.36$ TeV in the CMS collaboration 
 by Eq.(\ref{eq.int01}), where $\langle n_{\rm ch}\rangle$ is used instead of 
 $\langle n \rangle$. }
  \label{tab.cms2360FWb}
%  {\small
 \begin{center}
  \begin{tabular}{cccccc} \hline \hline
  $\sqrt{s}$ (TeV) & $k$ & $\Delta \eta$ &  $\tilde{p}$ & $\langle n_{\rm ch} \rangle$ 
                                                 & $\chi_{\rm min}^2/{\rm n.d.f.}$ \\ 
   \hline
 $2.36$&$1$ &$0.5$&$0.858\pm 0.022$  & $ 4.72\pm 0.09$ & $44.0/(23-2)$  \\ 
          &     &$1.0$& $0.782\pm 0.028$  & $ 8.88\pm 0.27$ & $326.8/(40-2)$ \\     
          &     &$1.5$& $0.810\pm 0.017$  & $14.2\pm 0.3   $ & $160.6/(50-2)$ \\     
          &     &$2.0$& $0.808\pm 0.015$  & $19.24\pm 0.3 $ & $165.5/(60-2)$ \\     
          &     &$2.4$& $0.795\pm 0.013$  & $23.51\pm 0.39$ & $157.6/(70-2)$ \\ 
  \hline    
 $2.36$&$2$&$0.5$& $1.000\pm 0.049$ & $ 4.90\pm 0.14$ & $124.0/(23-2)$  \\ 
         &     &$1.0$& $1.000\pm 0.032$ & $ 9.80\pm 0.22$ & $177.4/(40-2)$ \\     
         &     &$1.5$& $1.000\pm 0.026$ & $14.9 \pm 0.27$ & $155.2/(50-2)$ \\     
         &     &$2.0$& $1.000\pm 0.021$ & $19.9 \pm 0.3 $ & $137.8/(60-2)$ \\     
         &     &$2.4$& $1.000\pm 0.016$ & $23.9 \pm 0.3 $ & $ 98.8/(70-2)$ \\ 
  \hline \hline
   \end{tabular}  
 \end{center}
% }
 \end{table}

Comparing each value of $\chi_{\rm min}^2/{\rm n.d.f}$ in Table \ref{tab.cms0900FW} 
and that in Table \ref{tab.cms0900FWb}, the ratio of the former to the latter is from  
0.79 to 0.91 at $k=1$, and from 0.74 to 0.87 at $k=2$.

In the comparison of  each value of $\chi_{\rm min}^2/{\rm n.d.f}$ in Table \ref{tab.cms2360FW} 
and that in Table \ref{tab.cms2360FWb}, the ratio is from  
0.79 to 0.92 at $k=1$, and from 0.86 to 0.92 at $k=2$.

Therefore, fitting with Eqs.(\ref{eq.tl06}) and (\ref{eq.anal03}) becomes better than 
that with Eq.(\ref{eq.int01}).
 
Similar calculations with $k=1$ in Tables \ref{tab.cms0900FWb} and \ref{tab.cms2360FWb} were reported in \cite{Ghosh2011}.

%---------------------------
\section{Analysis by GL formula}
\label{apdx.GL}
%---------------------------

In order to compare the results by Eqs.(\ref{eq.tl06}) and (\ref{eq.anal03}), where charge conservation in the full phase space is taken into account, we also analyze the data by the GL formula, Eq.(\ref{eq.int05}) with two parameters, $p_{\rm in}$ and 
$\langle n_{\rm ch} \rangle$, which is used in place of $\langle n \rangle$.
Results at $\sqrt{s}=0.9$,  $2.36$ and $7$ TeV in the CMS Collaboration are shown  in Table \ref{tab.cms_GLkam}. 

In the comparison of  each value of $\chi_{\rm min}^2/{\rm n.d.f}$ in Table 
 \ref{tab.cms_GLkam} and that in Table \ref{tab.cms_GL}, the former is smaller than the latter.  Therefore, fitting with Eqs.(\ref{eq.tl06}) and (\ref{eq.anal03}) becomes better than that with Eq.(\ref{eq.int05}).

As can be seen from Tables \ref{tab.cms_GL} and \ref{tab.cms_GLkam}, 
$\langle n_{\rm ob} \rangle=2\zeta\langle n \rangle$ is almost the same with
 $\langle n_{\rm ch} \rangle$ in each analysis of tha data.

%---------- cms0.9-7_GL_pam ----------
%
 \begin{table}[htb]
  \caption{Parameters estimated from the analysis of charged multiplicity 
 distributions at $\sqrt{s}=0.9, 2.36$ and $7$ TeV in the CMS collaboration 
 by Eq.(\ref{eq.ap_ob12}) with k=1 and $\langle n_{\rm ch} \rangle$ in place of 
 $\langle  n \rangle$.  }
  \label{tab.cms_GLkam}
%  {\small
 \begin{center}
  \begin{tabular}{ccccc} \hline \hline
  $\sqrt{s}$ (TeV) & $\Delta \eta$ & $p_{\rm in}$ & $\langle n_{\rm ch} \rangle$ 
                                                 & $\chi_{\rm min}^2/{\rm n.d.f.}$ \\ 
 \hline
  $0.9$& $0.5$ & $0.441\pm 0.08$  & $ 3.62\pm 0.02$  & $ 3.1/(23-2)$ \\ 
          & $1.0$ & $0.374\pm 0.01$  & $ 7.29\pm 0.07$  & $ 37.6/(40-2)$ \\     
          & $1.5$ & $0.359\pm 0.09$  & $11.0 \pm 0.1  $  & $ 40.5/(52-2)$ \\     
          & $2.0$ & $0.334\pm 0.08$  & $14.9 \pm 0.1  $  & $ 46.5/(62-2)$ \\     
          & $2.4$ & $0.311\pm 0.07$  & $18.0 \pm 0.1  $  & $63.5/(68-2)$ \\ 
 \hline    
  $2.36$& $0.5$ & $0.495\pm 0.014$ & $4.62 \pm0.03$ & $ 7.4/(23-2)$ \\ 
           & $1.0$ & $0.418\pm 0.014$ & $9.22 \pm 0.11$ & $46.0/(40-2)$ \\     
           & $1.5$ & $0.430\pm 0.011$ & $14.0 \pm 0.1  $ & $26.8/(52-2)$ \\     
           & $2.0$ & $0.397\pm 0.012$ & $18.9 \pm 0.2  $ & $51.9/(62-2)$ \\     
           & $2.4$ & $0.369\pm 0.011$ & $22.9 \pm 0.2  $ & $55.3/(68-2)$ \\ 
 \hline         
     $7$ & $0.5$ & $0.595\pm 0.020$ & $5.98 \pm 0.06$ & $ 85.0/(23-2)$ \\ 
           & $1.0$ & $0.542\pm 0.012$ & $12.1 \pm 0.1  $ & $133.7/(40-2)$ \\     
           & $1.5$ & $0.501\pm 0.010$ & $18.4 \pm 0.2  $ & $195.8/(52-2)$ \\     
           & $2.0$ & $0.484\pm 0.009$ & $24.9 \pm 0.1 $ & $206.8/(62-2)$ \\     
           & $2.4$ & $0.495\pm 0.007$ & $30.3 \pm 0.1 $ & $135.1/(68-2)$ \\                             
  \hline \hline
   \end{tabular}  
 \end{center}
 %}
 \end{table} 
%
%---------------------------------------------------------

%
%\newpage
\hspace{5mm}

\end{document}